\documentclass[twocolumn,aps,showpacs,amsmath,amssymb]{revtex4}
\usepackage{graphics}
\usepackage{psfrag}

\begin{document}

\title{
  Renormalization of the periodic Anderson model:
  an alternative analytical approach to heavy Fermion behavior
}

\author{A. H\"{u}bsch$^{a,b}$ and K. W. Becker$^{a}$}
  \affiliation{
    $^{a}$Institut f\"{u}r Theoretische Physik,
    Technische Universit\"{a}t Dresden, D-01062 Dresden, Germany \\
    $^{b}$Department of Physics, University of California, Davis, CA 95616
  }

\date{\today}

\begin{abstract}
In this paper a recently developed projector-based renormalization method 
(PRM) for many-particle Hamiltonians is applied to the periodic Anderson model 
(PAM) with the aim to describe heavy Fermion behavior. In this method 
high-energetic excitation operators instead of high energetic states  
are eliminated. We arrive at an effective Hamiltonian for a quasi-free system 
which consists  of two non-interacting heavy-quasiparticle bands. The 
resulting renormalization equations for the parameters of the Hamiltonian are 
valid for large as well as small degeneracy $\nu_f$ of the angular momentum. 
An expansion in $1/\nu_f$ is avoided. Within an additional approximation which 
adapts the idea of a fixed renormalized \textit{f} level 
$\tilde{\varepsilon}_{f}$, we obtain coupled equations for 
$\tilde{\varepsilon}_{f}$ and the averaged \textit{f} occupation 
$\langle n_{f} \rangle$. These equations resemble to a certain extent those of 
the usual slave boson mean-field (SB) treatment. In particular, for large 
$\nu_f$ the results for the PRM and the SB approach agree perfectly whereas 
considerable differences are found for small $\nu_f$.
\end{abstract}


\maketitle

\begin{widetext}

\section{Introduction}

In comparison to ordinary metals metallic heavy fermion systems have 
remarkable low-temperature properties \cite{Ott}: both the
conduction electron specific heat and the magnetic susceptibility can be two
or more orders of magnitude larger than in normal 
metals though the ratio of both quantities is similar to   
that of usual metals. Usually, in metals an
increasing resistivity $\rho(T)$ with increasing temperature is observed. 
In contrast, a much richer behavior is found in
heavy fermion systems: At higher temperatures $\rho(T)$ only changes slightly 
and might even increase with decreasing temperature. Below a 
characteristic coherence temperature a strong decrease of the 
resistivity with decreasing
temperatures is observed. At very low temperatures, a $T^{2}$ dependence
of the temperature is found. Another important finding is 
that a correspondence between the low-energy excitations of 
heavy fermion systems and those of a free electron gas with properly 
renormalized parameters can be established. Therefore,
the high density of states at the Fermi surface observed in heavy fermion 
systems implies an effective mass of the (heavy) quasiparticles 
which is some hundred times larger than the free electron mass.

Prototype heavy fermion systems like CeAl$_{3}$ and  UPt$_{3}$ contain 
rare-earth or actinide elements. 
Thus, the basic microscopic model for the investigation of such materials 
is believed to be the periodic Anderson model (PAM) which describes the 
interaction between nearly localized, strongly correlated \textit{f} electrons 
and conduction electrons \cite{Lee}. Within a simplified version the 
Hamiltonian of the PAM can be written as
\begin{eqnarray}
  \label{G1}
  \mathcal{H} &=& \mathcal{H}_{0} + \mathcal{H}_{1} , \\[2ex]
  \mathcal{H}_0 & = & \varepsilon_{f} \sum _{i,m} \hat{f}^{\dagger} _{im}
  \hat{f} _{im}
  + \sum _{{\bf k},m} \varepsilon_{{\bf k}} \ c^{\dagger}_{{\bf k}m}
  c_{{\bf k}m} , \nonumber\\
  \mathcal{H}_1 & = & \frac{1}{\sqrt{N}} \sum _{{\bf k},i,m} V_{{\bf k}}
  \left(
    \hat{f}^{\dagger} _{im} c_{{\bf k}m} \,
    e^{{\rm i}{\bf k}{\bf R}_i} + {\rm h.c.}
  \right ) .
  \nonumber
\end{eqnarray}
Here, \textit{i} is the 4\textit{f} or 5\textit{f} 
site index, ${\bf k}$ is the conduction 
electron wave vector, and $V_{\bf k}$ is the hybridization matrix element
between conduction and localized electrons. $\varepsilon_{f}$ and
$\varepsilon_{\mathbf{k}}$, both measured from the chemical potential $\mu$,
are the excitation energies for $f$ and conduction electrons, respectively.
As a simplification, both types of electrons are assumed to have the same 
angular momentum index $m$ with $\nu_{f}$ values, $m=1 ... \nu_{f}$. Finally, 
the local Coulomb repulsion $U_{f}$ at $f$ sites has been assumed to be 
infinitely large so that localized sites can either be empty or singly 
occupied, i.e., the Hubbard operators 
${\hat f}_{im}^{\dagger}$ are defined by
\begin{eqnarray}
  \label{G2}
  \hat{f}^{\dagger} _{im} = f^{\dagger} _{im} \prod_{\tilde m (\ne m)}
  (1- n^f _{i \tilde m} )
\end{eqnarray}
where $n^f _{im}= f_{im}^\dagger f_{im}$.
The unexpected and exciting properties of the PAM \eqref{G1} are mainly due to 
the presence of  the strong correlations at \textit{f} sites. In turn the
strong correlations also cause the great difficulties in any theoretical
treatment of the model. In the present approach the correlations are 
taken care of by the Hubbard operators \eqref{G2} which do not obey the usual
fermionic anticommutator relations. Instead one has
\begin{eqnarray}
  \label{G18}
  [\hat{f}^{\dagger}_{im}, \hat{f}_{im}]_{+} &=& 
  {\cal D}_{im}
\end{eqnarray}
where
\begin{eqnarray*}
  {\cal D}_{im} &=&  
  \prod_{{\tilde m} ( \neq m) } 
  (1 - f_{i \tilde{m}}^{\dagger}f_{i \tilde{m}} ).
\end{eqnarray*}
The quantity ${\cal D}_{im}$ can be interpreted as a local projection operator 
at $f$ site $i$ on $f$ states which are either empty or singly occupied with 
one electron with index $m$. Also it is helpful to introduce separately the 
projection operator ${\cal P}_0(i)$ on the empty $f$ state at site $i$  and
the projection operator $\hat{n}_{im}^f$ on the singly occupied $f$ state when 
one electron with index $m$ is present. ${\cal D}_{im}$ can be rewritten
as 
\begin{eqnarray}
\label{G19}
  {\cal D}_{im} &=&  {\cal P}_0(i) + \hat{n}_{im}^f \,=\, 
  1 - \sum_{{\tilde m}(\neq m)} \hat{n}_{i {\tilde m}}^{f}
\end{eqnarray}
where we have defined
\begin{eqnarray}
  \label{G20}
  {\cal P}_0(i) &=& 
  \prod_{{\tilde m}} (1 - f_{i\tilde{m}}^{\dagger} f_{i\tilde{m}}),\\
  \label{G21}
  \hat{n}_{im}^{f} &=& \hat{f}_{im}^\dagger   \hat{f}_{im} \,=\,
  f_{im}^{\dagger} f_{im}\,\prod_{{\tilde m}(\neq m)} 
  (1 - f_{i\tilde{m}}^{\dagger} f_{i\tilde{m}}).
\end{eqnarray}
The second equation in \eqref{G19} is the completeness relation for 
$f$ electrons at site $i$.

\bigskip
For the case of vanishing Coulomb repulsion $U_{f}$ the PAM \eqref{G1} is 
equivalent to the Fano-Anderson model \cite{Anderson,Fano}
which can be easily solved (see, for example, appendix A). However, 
much of the physics of the correlated 
model can also be understood in terms of a renormalization 
of the parameters of the 
uncorrelated Fano-Anderson model. Various theoretical 
methods have been developed in the past  
to generate such renormalized Hamiltonians, for instance  
the Gutzwiller projection \cite{Rice} or the slave-boson mean-field (SB) 
theory \cite{Coleman,Fulde}. Here we use a  recently 
developed \cite{Becker} projector-based renormalization method (PRM) to map 
the PAM to a free system consisting of two bands of uncorrelated 
quasi-particles. Furthermore, we avoid an expansion with respect to the 
degeneracy $\nu_{f}$ of the angular momentum and take all $1/\nu_{f}$ 
corrections into account.

The PRM has already been applied before to the PAM in 
Ref.~\onlinecite{Becker}. However, in the present approach the treatment from 
Ref.~\onlinecite{Becker} will be  improved 
in various points: 
(i) The PRM is performed in a completely non-perturbative manner.
(ii) All $1/\nu_{f}$ corrections are taken into account.
(iii) The dispersion of both quasiparticle bands is considered. 

Furthermore, we shall compare the results of the PRM with 
those of the SB treatment in much more detail.

The paper is organized as follows. First, in Sec.~\ref{PRM} we briefly repeat 
the recently developed PRM \cite{Becker}. In Sec.~\ref{PAM} the PRM is applied 
to the PAM whereby the renormalization equations for the model parameters are 
derived non-perturbatively. An analytical solution of the  renormalization
equations is found in Sec.~\ref{Ana} using a constant renormalized \textit{f} 
level $\tilde{\varepsilon}_{f}$. Furthermore, we compare our results with 
the solutions of the SB theory. Finally, our conclusions are presented in 
Sec.~\ref{Conclusions}.


\section{Projector-based renormalization Method (PRM)}
\label{PRM}

The PRM \cite{Becker} starts from a decomposition of a given many-particle 
Hamiltonian $\mathcal{H}$ into an unperturbed part ${\cal H}_{0}$ and
into a perturbation ${\cal H}_{1}$, 
\begin{eqnarray}
  \label{G3}
  {\cal H} &=& {\cal H}_{0} + {\cal H}_{1}.
\end{eqnarray}
We assume that the eigenvalue problem of ${\cal H}_{0}$ is known
\begin{eqnarray}
  \label{G4}
  {\cal H}_{0} |n\rangle &=& E^{(0)}_n |n\rangle .
\end{eqnarray}
${\cal H}_{1}$ is the interaction. Its presence usually prevents the exact 
solution of
the eigenvalue problem of the full Hamiltonian. Let us define a projection 
operator ${\bf P}_{\lambda}$ by
\begin{eqnarray}
  \label{G5}
  {\bf P}_{\lambda}\mathcal{A} &=& 
  \sum_{
    \genfrac{}{}{0pt}{1}{
      \genfrac{}{}{0pt}{1}{m,n}{
        \left| E_n^{(0)} - E_m^{(0)} \right| \leq \lambda
      }
    }{}
  }
  |n\rangle \langle m| \, \langle n| \mathcal{A}  |m\rangle .
\end{eqnarray}
Note that ${\bf P}_{\lambda}$ is a super-operator acting on usual
operators $\mathcal{A}$ of the unitary space. It projects on those parts 
of $\mathcal{A}$ which are
formed by all dyads $|n\rangle \langle m|$ with energy differences
$|E_n^{(0)} - E_m^{(0)}|$ less or equal to a given cutoff $\lambda$,
where $\lambda$ is  smaller than the cutoff $\Lambda$ of the original model. 
Note that in \eqref{G5} neither  $|n\rangle$ nor $|m\rangle$ have to be 
low-energy eigenstates of $\mathcal{H}_{0}$. However, their energy difference 
has to be restricted to values $\leq \lambda$. Furthermore, it is useful to 
define the projection operator
\begin{eqnarray}
  \label{G6}
  {\bf Q}_{\lambda} &=& {\bf 1} - {\bf P}_{\lambda}
\end{eqnarray}
which is orthogonal to ${\bf P}_{\lambda}$. ${\bf Q}_{\lambda}$ projects on
high energy transitions larger than the cutoff $\lambda$.

The goal of the present method is to transform the initial Hamiltonian
${\cal H}$ (with a large energy cutoff $\Lambda$) into an effective Hamiltonian
${\cal H}_{\lambda}$ which has no matrix elements belonging to transitions
larger than $\lambda$. This is achieved by an unitary transformation so that 
the effective Hamiltonian will have the same eigenspectrum as the original 
Hamiltonian ${\cal H}$. However, as it will turn out, the method 
is especially suitable
to describe low-energy excitations of the system. ${\cal H}_{\lambda}$ is 
defined by
\begin{eqnarray}
  \label{G7}
  {\cal H}_{\lambda} &=&
  e^{X_{\lambda}} \, {\cal H} \, e^{-X_{\lambda}} .
\end{eqnarray}
The generator $X_{\lambda}$ of the transformation has to be anti-Hermitian, 
$X^{\dagger}_{\lambda}=-X_{\lambda}$, so that ${\cal H}_{\lambda}$ is 
Hermitian for any $\lambda$. We look for an appropriate  generator 
$X_{\lambda}$ so that ${\cal H}_{\lambda}$ has no matrix elements belonging to 
transitions larger than $\lambda$. This means that the following condition
\begin{eqnarray}
  \label{G8}
  {\bf Q}_{\lambda}{\cal H}_{\lambda} &=& 0
\end{eqnarray}
has to be fulfilled. Eq.~\eqref{G8} will be used below 
to specify $X_{\lambda}$. In 
contrast to Ref.~\cite{Becker}, where $\mathcal{H}_{\lambda}$ 
was evaluated perturbatively, the transformation \eqref{G7} 
will be treated non-perturbatively. 

Next we discuss the elimination procedure for the interaction 
$\mathcal{H}_{1}$. Instead of transforming the Hamiltonian in one step 
as in Eq.~\eqref{G7} the transformation will be done successively. 
Or more formally 
spoken, instead of applying the elimination of high-energy excitations in one 
step a sequence of stepwise transformations is used in order to obtain an 
effectively diagonal model. This procedure resembles Wegner's flow equation 
method \cite{Wegner} and the similarity renormalization \cite{Glazek} in some 
aspects. In the PRM approach difference equations for the $\lambda$
dependence of the parameters of the Hamiltonian are derived. 
They will be called renormalization equations. To find these equations we
start from the renormalized Hamiltonian 
\begin{eqnarray}
  \label{G9}
  \mathcal{H}_{\lambda} &=& \mathcal{H}_{0,\lambda} + \mathcal{H}_{1,\lambda}
\end{eqnarray}
after all excitations with energy differences larger than $\lambda$ have been 
eliminated. Next we consider an additional renormalization of 
$\mathcal{H}_{\lambda}$ by eliminating all excitations inside an energy shell 
between $\lambda$ and a smaller energy cutoff $(\lambda - \Delta\lambda)$ 
where $\Delta\lambda > 0$. The new Hamiltonian 
$\mathcal{H}_{(\lambda-\Delta\lambda)}$ is obtained by an unitary 
transformation similar to that of Eq.~\eqref{G7}
\begin{eqnarray}
  \label{G10}
  \mathcal{H}_{(\lambda-\Delta\lambda)} &=& 
  e^{X_{\lambda,\Delta\lambda}} 
  \, \mathcal{H}_{\lambda} \, 
  e^{-X_{\lambda,\Delta\lambda}} 
\end{eqnarray}
where $X_{\lambda,\Delta\lambda}$ is determined by 
\begin{eqnarray}
  \label{G11}
  {\bf Q}_{(\lambda-\Delta\lambda)} {\cal H}_{(\lambda-\Delta\lambda)} &=& 0.
\end{eqnarray}
Note that there are two strategies to exploit Eq.~\eqref{G11} in order to
determine the generator $X_{\lambda,\Delta\lambda}$ of the unitary
transformation \eqref{G10}. The most straightforward route is to analyze
Eqs.~\eqref{G10} and \eqref{G11} in perturbation theory as it was done in
Refs.~\onlinecite{Becker} and \onlinecite{Huebsch}. Here, we want to perform
the renormalization step from $\lambda$ to $(\lambda - \Delta\lambda)$ in a
non-perturbative way.

Eqs.~\eqref{G10} and \eqref{G11} describe the renormalization 
of the Hamiltonian by decreasing the cutoff from $\lambda$ to 
$(\lambda - \Delta \lambda)$ and can be 
used to derive difference equations for the $\lambda$-dependence 
of the Hamiltonian. The 
resulting equations for the parameters of the Hamiltonian will be  
called renormalization equations. Their solution depends 
on the initial values of the parameters of
the Hamiltonian and fixes the final Hamiltonian in the limit 
$\lambda \rightarrow 0 $. Note that for $\lambda \rightarrow 0$ the resulting 
Hamiltonian only consists of the unperturbed part 
${\cal H}_{0,(\lambda \rightarrow 0)}$. 
The interaction ${\cal H}_{1,(\lambda\rightarrow 0)}$ 
vanishes since it is completely used up in the renormalization 
procedure. Thus, an effectively diagonal Hamiltonian is obtained.


\section{Renormalization of the periodic Anderson model (PAM)}
\label{PAM}

The PRM described above will be applied in this section to the PAM \eqref{G1}.
As an illustration of the method the Fano-Anderson model is discussed as a
further application of the PRM in appendix \ref{Fano}. This model can be
considered as a PAM without electronic correlations. It turns out that the
renormalization of the full PAM \eqref{G1} is somewhat similar to that of the
uncorrelated model. However, in the Anderson-Fano model the elimination of
excitations with energies larger than $\lambda$ can be done in one step. For
the PAM \eqref{G1} the $f$ electron one-particle energy $\varepsilon_f$ will
also be renormalized. This is due to the presence of strong local correlations
at $f$ sites in the PAM \eqref{G1}. Therefore, the elimination procedure has
to be done stepwise by repeatedly integrating over small energy steps of width 
$\Delta \lambda$. In this way one is led to renormalization equations for the 
parameters of the model in terms of difference equations which have to be 
solved.

\subsection{Renormalization ansatz}
Let us start by formally writing down the effective Hamiltonian 
${\cal H}_\lambda=e^{X_\lambda} {\cal H} e^{-X_\lambda}= {\cal H}_{0,\lambda}
+ {\cal H}_{1,\lambda}$  for the periodic Anderson model
after all excitations with energy differences larger than $\lambda$ have been 
eliminated. By comparing with the starting model \eqref{G1} 
one might be in favor of choosing an unperturbed part 
${\cal H}_{0,\lambda}$ which contains correlated $f$-electrons
$\hat{f}_{im}^\dagger, \hat{f}_{im}$ as in Eq.~\eqref{G1}. 
However, because the eigenvalue problem of such an 
Hamiltonian would not exactly be solvable, we prefer to  start 
from an uncorrelated Hamiltonian where only usual Fermi
operators $f_{im}^\dagger, f_{im}$  enter  
but keep the correlations in the renormalized 
interaction ${\cal H}_{1,\lambda}$, i.e.
\begin{eqnarray*}
  {\cal H}_{\lambda} &=&
  {\cal H}_{0,\lambda} + {\cal H}_{1,\lambda}
\end{eqnarray*}
where
\begin{eqnarray}
  \label{G12}
  {\cal H}_{0,\lambda} &=&
  \varepsilon_{f,\lambda}
  \sum_{\mathbf{k},m} f^{\dagger}_{\mathbf{k}m} f_{\mathbf{k}m}
  + \sum_{\mathbf{k},m} 
  \Delta_{\mathbf{k}, \lambda}\
  \left(
    f_{\mathbf{k}m}^{\dagger} f_{\mathbf{k}m} 
  \right)_{\mathrm{NL}}
  + \sum_{{\bf k},m}
  \varepsilon_{{\bf k},\lambda} \, c^{\dagger}_{{\bf k}m} c_{{\bf k}m} 
  + E_{\lambda} , \\[2ex]
  \label{G13}
  {\cal H}_{1,\lambda} &=& {\bf P}_{\lambda}{\cal H}_{1} \,=\,
  \sum_{\mathbf{k},m}
  V_{\mathbf{\bf k}} \,
  {\bf P}_\lambda \left(
    \hat{f}^{\dagger}_{\mathbf{k}m} c_{{\bf k}m} \, + \mathrm{h.c.}
  \right).
\end{eqnarray}
Here $f^{\dagger}_{\mathbf{k}m}, \hat{f}^\dagger_{\mathbf{k}m}$ are
Fourier transformed \textit{f} operators,
\begin{eqnarray}
  \label{G14}
  f^{\dagger}_{\mathbf{k}m} &=& 
  \frac{1}{\sqrt{N}} \sum_{i} 
  f^{\dagger}_{im} \, e^{i \mathbf{k}\mathbf{R}_{i}}, \qquad
  \hat{f}^{\dagger}_{\mathbf{k}m} \,=\, 
  \frac{1}{\sqrt{N}} \sum_{i} 
  \hat{f}^{\dagger}_{im} \, e^{i \mathbf{k}\mathbf{R}_{i}}.
\end{eqnarray}
Moreover, in \eqref{G12} we have discriminated 
between local, 
\begin{eqnarray}
  \label{G16}
  \left( f^{\dagger}_{m} f_{m} \right)_{\mathrm{L}} &:=&
  \frac{1}{N} \sum_{\mathbf{k}} 
  f_{\mathbf{k}m}^{\dagger} f_{\mathbf{k}m} \,=\,
  \frac{1}{N} \sum_{i} f_{im} ^{\dagger} f_{im}, 
  \end{eqnarray}
and nonlocal,
\begin{eqnarray}
  \label{G17}
  \left(
    f_{\mathbf{k}m}^{\dagger} f_{\mathbf{k}m} 
  \right)_{\mathrm{NL}}  
  &:=&
  \frac{1}{N} \sum_{i,j(\neq i)} f^{\dagger}_{im} f_{jm} \,
  e^{i{\bf k}({\bf R}_i -{\bf R}_j)}
  \,=\,
  f_{\mathbf{k}m}^{\dagger} f_{\mathbf{k}m} -
  \left( f^{\dagger}_{m} f_{m} \right)_{\mathrm{L}}, 
  \end{eqnarray}
$f$ particle-hole excitations in order to properly take into 
account the strong Coulomb interaction at local $f$ sites.
Due to renormalization processes the one-particle energies
$\varepsilon_{f,\lambda}$ and $\varepsilon_{\mathbf{k},\lambda}$ in 
Eq.~\eqref{G12} depend on the cutoff energy  $\lambda$. Moreover, two new 
parameters enter: $\Delta_{\mathbf{k},\lambda}$ describes the \textit{f} 
dispersion due to the hybridization of  $f$ electrons at different sites
$i \neq j$, and  $E_\lambda$ is an additional energy shift.
Finally, the projector ${\bf P}_{\lambda}$ in \eqref{G13} guarantees that only 
excitations contribute to ${\cal H}_{1,\lambda}$ which have
energies (with respect to ${\cal H}_{0,\lambda}$) which are smaller
than $\lambda$. The initial parameter values of the original model (at 
$\lambda = \Lambda$) are
\begin{eqnarray}
  \label{G15}
  \varepsilon_{f,(\lambda= \Lambda)} \,=\, \varepsilon_f, 
  \qquad
  \Delta_{\mathbf{k},(\lambda= \Lambda)} \,=\,0, 
  \qquad
  \varepsilon_{{\bf k}, (\lambda= \Lambda)} \,=\, \varepsilon_{\bf k},
  \qquad
  E_{(\lambda= \lambda)} \,=\, 0.
\end{eqnarray}
As it turns out, the hybridization $V_{\mathbf{k}}$ is not changed by the
renormalization procedure. 

\bigskip
As mentioned before, correlation effects have been
neglected in Eq.~\eqref{G12}. First, this means 
that doubly occupancies of $f$-sites 
$(f_{im}^{\dagger} - \hat{f}_{im}^{\dagger})(f_{im} - \hat{f}_{im})$
are assumed to be negligibly small though they are not properly 
excluded by the choice of uncorrelated $f$ operators in 
${\cal H}_{0,\lambda}$. Note that this assumption is also used 
for the subsidiary condition within the SB approach \cite{Fulde}. 
Doubly occupied $f$-sites could in principle be generated 
by the non-local $f$-part of  ${\cal H}_{0,\lambda}$. As it turns out, 
this is explicitely excluded by keeping 
the correlations in the interaction part ${\cal H}_{1,\lambda}$. 

\bigskip
For the following, the commutator of the unperturbed Hamiltonian 
${\cal H}_{0,\lambda}$ with the operator 
$\hat{f}_{k,m}^\dagger c_{km}$ has to be 
evaluated. By introducing the unperturbed Liouville operator 
${\bf L}_{0,\lambda}$, which is defined by 
$\mathbf{L}_{0,\lambda}\mathcal{A} = [\mathcal{H}_{0,\lambda},\mathcal{A}]$
for any operator variable $\mathcal{A}$, one finds
\begin{eqnarray}
  \label{G22}
  \lefteqn{
    {\bf L}_{0,\lambda} \hat{f}^{\dagger}_{\mathbf{k}m} c_{\mathbf{k}m} 
    \,=\,
    [ {\cal H}_{0,\lambda},  \hat{f}^{\dagger}_{\mathbf{k}m} c_{\mathbf{k}m} ] 
    \,=\,
  }\\
  &=& 
  \left(
    \varepsilon_{f,\lambda} - \varepsilon_{\mathbf{k},\lambda} 
  \right) \,
  \hat{f}^{\dagger}_{\mathbf{k}m} c_{\mathbf{k}m} 
  +
  \frac{1}{N^{3/2}} \sum_{\mathbf{k}',i,j} (1 - \delta_{ij}) 
  \Delta_{\mathbf{k}', \lambda} \,
  e^{i({\bf k} -{\bf k}'){\bf R}_j} \, e^{i{\bf k}'{\bf R}_i}
  f_{im}^\dagger {\cal D}_{jm} c_{\mathbf{k}m} 
  \nonumber \\
  &&
  +
  \frac{1}{N^{3/2}} \sum_{\mathbf{k}',m',i,j} (1 - \delta_{mm'})
  \Delta_{\mathbf{k}', \lambda} \,
  e^{i({\bf k} -{\bf k}'){\bf R}_j} \, e^{i{\bf k}'{\bf R}_i}
  f_{im}^\dagger \hat{f}_{jm}^{\dagger} \hat{f}_{jm'} c_{\mathbf{k}m} 
  \nonumber
\end{eqnarray}
where contributions which lead to doubly occupied $f$ sites have been
neglected. The second and the third term on the r.h.s. of Eq.~\eqref{G22}
follow from the special form of the anticommutator relations
\eqref{G18}. Obviously, only $f$ electron operators belonging to different
sites $i \not= j$ enter the second term on the r.h.s. of
Eq.~\eqref{G22}. Therefore, as approximation one may replace the operator 
$\mathcal{D}_{jm}$ by its expectation value
\begin{eqnarray}
  \label{G23}
  D &=& \langle \mathcal{D}_{jm} \rangle \,=\,
  1 - \frac{\nu_{f} - 1}{\nu_{f}} \langle \hat{n}_{j}^{f} \rangle
\end{eqnarray}
where
\begin{eqnarray}
  \label{G24}
  \langle \hat{n}_{j}^{f} \rangle &=& 
  \sum_{m} \left\langle
    \hat{f}_{jm}^{\dagger} \hat{f}_{jm} 
  \right\rangle
\end{eqnarray}
is the averaged occupation number of $f$ electrons at site $j$. Note that $D$ 
is independent of $j$ and $m$. Furthermore, $D \, f_{\mathbf{k}m}^{\dagger}$
is replaced by $\hat{f}_{\mathbf{k}m}^{\dagger}$. Finally, 
we neglect the third term on the r.h.s of Eq.~\eqref{G22}, which represents 
spin-flip processes, and all contributions leading to doubly occupied $f$ 
sites. (Similar approximations will also be used later.) Consequently, 
Eq.~\eqref{G22} simplifies to
\begin{eqnarray}
  \label{G25}
  {\bf L}_{0,\lambda} \hat{f}^{\dagger}_{\mathbf{k}m} c_{\mathbf{k}m} &=&
  \left(
    \varepsilon_{f,\lambda} + \Delta_{\mathbf{k},\lambda} - 
    \bar{\Delta}_{\lambda} - \varepsilon_{\mathbf{k},\lambda} 
  \right) \,
  \hat{f}^{\dagger}_{\mathbf{k}m} c_{\mathbf{k}m}
\end{eqnarray}
where
\begin{eqnarray}
  \label{G26}
  \bar{\Delta}_{\lambda} &=& 
  \frac{1}{N} \sum_{\mathbf{k}} \Delta_{\mathbf{k},\lambda}
\end{eqnarray}
is the averaged $f$ dispersion. Thus, 
$\hat{f}^{\dagger}_{\mathbf{k}m} c_{\mathbf{k}m}$ is an approximate
eigenvector of the Liouville operator ${\bf L}_{0,\lambda}$. The corresponding 
eigenvalue is the excitation energy 
$
  \varepsilon_{f,\lambda} + \Delta_{\mathbf{k},\lambda} - 
  \bar{\Delta}_{\lambda} - \varepsilon_{\mathbf{k},\lambda}
$.
Furthermore, Eq.~\eqref{G25} can be used to evaluate the 
action of the projector $\mathbf{P}_{\lambda}$ in \eqref{G13} so that  
${\cal H}_{1,\lambda}$ can be rewritten as
\begin{eqnarray}
  \label{G27}
  {\cal H}_{1,\lambda} &=& {\bf P}_{\lambda}{\cal H}_{1} \,=\,
  \sum_{\mathbf{k},m}
  \Theta\left( 
    \lambda - |\varepsilon_{f,\lambda} + \Delta_{\mathbf{k},\lambda} -
    \bar{\Delta}_{\lambda} - \varepsilon_{\mathbf{k},\lambda} | 
  \right) \,
  V_{\mathbf{k}}
  \left(
    \hat{f}^{\dagger}_{\mathbf{k}m} c_{\mathbf{k}m} \, + {\rm h.c.}
  \right)
\end{eqnarray}
where the $\Theta$-function restricts the particle-hole excitations
to transition energies smaller than $\lambda$.

\subsection{Generator of the unitary transformation}
In the next step let us evaluate a new effective Hamiltonian
${\cal H}_{(\lambda - \Delta \lambda)}$ which is obtained by a further 
elimination of excitations within a small energy shell between
$(\lambda-\Delta\lambda)$ and $\lambda$. According to \eqref{G10} 
${\cal H}_{(\lambda-\Delta\lambda)}$ is obtained from an unitary 
transformation 
\begin{eqnarray}
  \label{G28}
  \mathcal{H}_{(\lambda-\Delta\lambda)} &=& 
  e^{X_{\lambda,\Delta\lambda}} 
  \, \mathcal{H}_{\lambda} \, 
  e^{-X_{\lambda,\Delta\lambda}} 
\end{eqnarray}
where $X_{\lambda,\Delta\lambda}$ is the generator of the unitary 
transformation. 
For the explicit form of $X_{\lambda, \Delta \lambda}$
let us  make the following ansatz
\begin{eqnarray}
  \label{G29}
  X_{\lambda, \Delta \lambda} &=& 
  \sum_{\mathbf{k},m} 
  A_{\mathbf{k}}(\lambda, \Delta\lambda) \, 
  \Theta_{\mathbf{k}} (\lambda ,\Delta\lambda) \,
  ( \hat{f}_{\mathbf{k}m}^{\dagger} c_{\mathbf{k}m} - 
  c_{\mathbf{k}m}^{\dagger} \hat{f}_{\mathbf{k}m})
\end{eqnarray}
where $\Theta_{\mathbf{k}}(\lambda, \Delta\lambda)$ is the product of two
$\Theta$-functions
\begin{eqnarray}
  \label{G30}
  \Theta_{\mathbf{k}}(\lambda, \Delta\lambda) &=& 
  \Theta(
    \lambda - 
    |
      \varepsilon_{f,\lambda} + \Delta_{\mathbf{k},\lambda} - 
      \bar{\Delta}_{\lambda} -
      \varepsilon_{\mathbf{k},\lambda}
    |
  ) \\
  && \times
  \Theta\left[
    |
      \varepsilon_{f,(\lambda - \Delta\lambda)} +
      \Delta_{\mathbf{k},(\lambda - \Delta\lambda)} -
      \bar{\Delta}_{(\lambda - \Delta\lambda)} -
      \varepsilon_{\mathbf{k},(\lambda - \Delta\lambda)} 
    | - 
    (\lambda - \Delta\lambda)
    \right]
    \nonumber.
\end{eqnarray}
The operator form of $X_{\lambda, \Delta \lambda}$ is suggested by its
first order expression which is easily obtained by expanding \eqref{G10} in
powers of ${\cal H}_1$ and using  Eq.~\eqref{G11} (compare
Ref.~\onlinecite{Becker}). The yet unknown prefactors 
$A_{\mathbf{k}}(\lambda,\Delta\lambda)$ will be specified later and depend on 
$\lambda$ and $\Delta\lambda$. It will turn out that 
$A_{\mathbf k}(\lambda, \Delta\lambda)$ contains contributions to
all powers in $V_{\mathbf{k}}$. Note that the ansatz \eqref{G28}  also 
corresponds to the operator structure of the exact generator of the 
uncorrelated Fano-Anderson model (see appendix A). We expect the ansatz 
\eqref{G29} to be a good approximation also for the correlated model in which
conduction and localized $f$ electrons strongly couple. 
Finally, the $\Theta$-functions in \eqref{G29} result from the restriction of  
${\cal H}_\lambda$ to particle-hole excitations  with 
$
  |
    \varepsilon_{f,\lambda} + \Delta_{\mathbf{k},\lambda} - 
    \bar{\Delta}_{\lambda} -
    \varepsilon_{\mathbf{k},\lambda} 
  | < 
  \lambda
$ 
and from the corresponding restriction  
$
  |
    \varepsilon_{f,(\lambda - \Delta \lambda)} + 
    \Delta_{\mathbf{k},(\lambda - \Delta \lambda)} - 
    \bar{\Delta}_{(\lambda - \Delta\lambda)} -
    \varepsilon_{\mathbf{k},(\lambda - \Delta \lambda)} 
  | > 
  \lambda - \Delta \lambda
$
for the renormalized model ${\cal H}_{(\lambda - \Delta \lambda)}$. The two 
$\Theta$-functions in $\Theta_{\mathbf{k}}(\lambda, \Delta  \lambda)$  
confine the allowed excitations.

To determine the unknown parameters $A_{\mathbf{k}}(\lambda,\Delta\lambda)$ of
the unitary transformation [compare \eqref{G29}] we will use 
Eq.~\eqref{G11}. First, we have to carry out the unitary transformation
\eqref{G28} explicitly
\begin{eqnarray}
  \label{G31}
  {\cal H}_{(\lambda -\Delta \lambda)} &=&
  \varepsilon_{f,\lambda}
  \sum_{\mathbf{k},m}
  e^{X_{\lambda ,\Delta \lambda}}
  f^{\dagger}_{\mathbf{k}m} f_{\mathbf{k}m}
  e^{-X_{\lambda, \Delta \lambda}} + 
  \sum_{\mathbf{k},m} \Delta_{\mathbf{k}, \lambda} \,
  e^{X_{\lambda, \Delta \lambda}}
  \left( f_{\mathbf{k}m}^{\dagger} f_{\mathbf{k}m} \right )_{NL}\, 
  e^{-X_{\lambda, \Delta \lambda}} \\[1ex]
  && +
  \sum_{\mathbf{k},m} \varepsilon_{{\bf k},\lambda} \,
  e^{X_{\lambda, \Delta \lambda}}
  c^{\dagger}_{{\bf k}m} c_{{\bf k}m} 
  e^{- X_{\lambda, \Delta \lambda}}  +
  \sum_{{{\bf k},m}} V_{{\bf k}} \,
  e^{X_{\lambda, \Delta \lambda}}
  \left(
    \hat{f}^{\dagger}_{\mathbf{k}m} c_{{\bf k}m}  + \, {\rm h.c.} 
  \right)
  e^{-X_{\lambda, \Delta \lambda}} + E_{\lambda} \, . \nonumber
\end{eqnarray}
The transformations for the various operators are given in appendix
\ref{Transformation} [see Eqs. \eqref{B30}-\eqref{B33}]. For instance, the 
transformation of $c^{\dagger}_{{\bf k}m} c_{{\bf k}m}$ reads,
\eqref{B30}, 
\begin{eqnarray}
  \label{G32}
  \lefteqn{
    e^{X_{\lambda, \Delta \lambda}}
    c^{\dagger}_{{\bf k}m} c_{{\bf k}m} 
    e^{- X_{\lambda, \Delta \lambda}} -
    c^{\dagger}_{{\bf k}m} c_{{\bf k}m} 
    \,=\,
  }
  && \\[1ex]
  &=&
  - \,
  \frac{1}{2D} \, \Theta_{\mathbf{k}}(\lambda,\Delta\lambda) \,
  A_{\mathbf{k}}(\lambda, \Delta\lambda) \,
  \left\langle
    \hat{f}_{\mathbf{k}m}^{\dagger}c_{\mathbf{k}m}+\mathrm{h.c.}
  \right\rangle 
  \left[
    1 - D - 
    \sum_{\tilde{m}(\not=m)} 
    \left( f_{\tilde{m}}^{\dagger} f_{\tilde{m}} \right)_{\mathrm{L}}
  \right]
  \nonumber\\[1ex]
  &&
  -\,
  \frac{1}{2D} \, \Theta_{\mathbf{k}}(\lambda,\Delta\lambda) \,
  \left\{
    \cos \left[ 2\sqrt{D} A_{\mathbf{k}}(\lambda, \Delta\lambda) \right] - 
    1
  \right\}
  \left\{
    D \left( f_{\mathbf{k}m}^{\dagger} f_{\mathbf{k}m} \right)_{\mathrm{NL}} +
    \left( f_{m}^{\dagger} f_{m} \right)_{\mathrm{L}} 
    \phantom{\sum_{\tilde{m}(\not=m)}}
  \right. \nonumber \\[1ex]
  &&
  \left.
    \qquad\qquad -\,
    D c_{\mathbf{k}m}^{\dagger} c_{\mathbf{k}m} - 
    \left\langle c_{\mathbf{k}m}^{\dagger} c_{\mathbf{k}m} \right\rangle
    \left[
      1 - D - 
      \sum_{\tilde{m}(\not=m)} 
      \left( f_{\tilde{m}}^{\dagger} f_{\tilde{m}} \right)_{\mathrm{L}}
    \right]
  \right\} \nonumber \\[1ex]
  &&
  +\,
  \frac{1}{2\sqrt{D}} \, \Theta_{\mathbf{k}}(\lambda,\Delta\lambda) \,
  \sin \left[ 2\sqrt{D} A_{\mathbf{k}}(\lambda, \Delta\lambda) \right]
  \left\{
    \left( 
      \hat{f}_{\mathbf{k}m}^{\dagger}c_{\mathbf{k}m}+\mathrm{h.c.}
    \right) 
    \phantom{\sum_{\tilde{m}(\not=m)}}
  \right. \nonumber \\[1ex]
  &&
  \left.
    \qquad\qquad +\,
    \frac{1}{2D}
    \left\langle
      \hat{f}_{\mathbf{k}m}^{\dagger}c_{\mathbf{k}m}+\mathrm{h.c.}
    \right\rangle
    \left[
      1 - D - 
      \sum_{\tilde{m}(\not=m)} 
      \left( f_{\tilde{m}}^{\dagger} f_{\tilde{m}} \right)_{\mathrm{L}}
    \right]
  \right\}. \nonumber
\end{eqnarray}
Similar expressions can also be found for the 
transformation of the remaining operators. In deriving
\eqref{B30} - \eqref{B33} an additional  factorization approximation was used 
in order to keep only operators which are bilinear in the fermionic creation 
and annihilation operators. Spin-flip contributions have been neglected.
Moreover, it was assumed that the number of $\mathbf{k}$ points which are 
integrated out by the transformation from $\lambda$ to 
$(\lambda -\Delta \lambda)$ is small compared to the total number of
$\mathbf{k}$ points.

As already mentioned in the introduction, an expansion with respect to the
degeneracy $\nu_{f}$ is utilized in the slave-boson mean-field (SB)
theory \cite{Coleman, Fulde}. In contrast, here we incorporate all
$1/\nu_{f}$ corrections which will be reflected by the expectation value
$D$ as defined in Eq.~\eqref{G23}. As one can see from the two terms of the
anticommutator of Eq.~\eqref{G18}, new renormalization
contributions are included by which a localized electron at an occupied $f$
site is annihilated and instead a conduction electron is created. These
processes are of order $1/\nu_{f}$ smaller than the usual processes included
in the SB theory by which a conduction electron is annihilated and instead a
localized electron is generated at a formerly empty $f$ site.

In the next step let us  determine the unknown parameters 
$A_{\mathbf{k}}(\lambda, \Delta\lambda)$. For that purpose, we insert
$\mathcal{H}_{(\lambda-\Delta\lambda)}$ from \eqref{G31} into 
Eq.~\eqref{G11} and use the transformed quantities \eqref{B30} -
\eqref{B33}. Thus, one finds
\begin{eqnarray}
  \label{G33}
  \Theta_{\mathbf{k}}(\lambda,\Delta\lambda) 
  \tan \left[ 2\sqrt{D} A_{\mathbf{k}}(\lambda, \Delta\lambda) \right]
  &=&
  \Theta_{\mathbf{k}}(\lambda,\Delta\lambda) 
  \frac{
    2 \sqrt{D} V_{\mathbf{k}}
  }{
    \varepsilon_{f,\lambda} + 
    D \left( \Delta_{\mathbf{k},\lambda} - \bar{\Delta}_{\lambda} \right) - 
    \varepsilon_{\mathbf{k},\lambda}
  }.
\end{eqnarray}
The condition \eqref{G33} for $A_{\mathbf{k}}(\lambda, \Delta\lambda)$ 
guarantees that $\mathcal{H}_{(\lambda-\Delta\lambda)}$ does not contain 
matrix elements with transition energies larger than 
$(\lambda-\Delta\lambda)$. Obviously, there is a strong similarity between 
\eqref{G33} and the corresponding result \eqref{A11} of the Fano-Anderson 
model. However, the generator \eqref{G33} of the PAM contains some
deviations which reflect the influence of the strong electronic correlations
at \textit{f} sites. It is important to note that the
expression  \eqref{G33} for $A_{\mathbf k}(\lambda, \Delta\lambda)$ is 
non-perturbatively in $V_{\mathbf k}$ and is not restricted to some low order 
in $V_{\bf k}$. Moreover note that the values of 
$A_{\mathbf{k}}(\lambda, \Delta\lambda)$ are determined by Eq.~\eqref{G32} 
only for the case that the excitation energy 
$
  \left(
    \varepsilon_{f,\lambda} + \Delta_{\mathbf{k},\lambda} - 
    \bar{\Delta}_{\lambda} - \varepsilon_{\mathbf{k},\lambda} 
  \right)
$ 
fits in the energy shell restricted by 
$\Theta_{\mathbf{k}}(\lambda,\Delta\lambda)$. For all other 
excitations $A_{\mathbf{k}}(\lambda, \Delta\lambda)$ can be set equal to
zero. Thus, the parameter $A_{\mathbf{k}}(\lambda, \Delta\lambda)$ of the
generator $X_{\lambda, \Delta \lambda}$ is  given by
\begin{eqnarray}
  \label{G34}
  A_{\mathbf{k}}(\lambda, \Delta\lambda) &=& 
  \left\{ \begin{array}{*{2}{l}}
    \frac{\displaystyle 1}{\displaystyle 2\sqrt{D}} 
    \, \arctan \left[
      \frac{
        \displaystyle 2 \sqrt{D} V_{\mathbf{k}}
      }{
        \displaystyle \varepsilon_{f,\lambda} + 
        D \left( \Delta_{\mathbf{k},\lambda} - \bar{\Delta}_{\lambda} \right) -
        \varepsilon_{\mathbf{k},\lambda}
      }
    \right] \,
    &  \quad  \mbox{for} \, 
    \Theta_{\mathbf{k}}(\lambda,\Delta\lambda) = 1 \\[1ex]
    0 \,  & \quad  \mbox{for} \, 
    \Theta_{\mathbf{k}}(\lambda,\Delta\lambda) = 0
  \end{array}\right.
\end{eqnarray}

\subsection{Renormalization equations}\label{ren_equation}
In the following we derive the renormalization equations for the parameters 
of the Hamiltonian. For that purpose we compare two different expressions for 
$\mathcal{H}_{(\lambda-\Delta\lambda)}$. The first one is obtained by
rewriting the renormalization ansatz [Eqs.~\eqref{G12} and \eqref{G13}] at
cutoff $(\lambda-\Delta\lambda)$ 
\begin{eqnarray}
  \label{G35}
  {\cal H}_{(\lambda-\Delta\lambda)} &=&
  \varepsilon_{f,(\lambda-\Delta\lambda)}
  \sum_{\mathbf{k},m}
  f^{\dagger}_{\mathbf{k}m} f_{\mathbf{k}m}
  + \sum_{\mathbf{k},m} 
  \Delta_{\mathbf{k}, (\lambda-\Delta\lambda)}\
  \left( f_{\mathbf{k}m}^{\dagger} f_{\mathbf{k}m} \right)_{\mathrm{NL}}\\[1ex]
  &&
  + \,
  \sum_{{\bf k},m}
  \varepsilon_{{\bf k},(\lambda-\Delta\lambda)} \, 
  c^{\dagger}_{{\bf k}m} c_{{\bf k}m} +
  E_{(\lambda-\Delta\lambda)} + 
  \sum_{\mathbf{k},m}
  V_{\mathbf{\bf k}} \,
  {\bf P}_\lambda \left(
    \hat{f}^{\dagger}_{\mathbf{k}m} c_{{\bf k}m} \, + \mathrm{h.c.}
  \right).
  \nonumber
\end{eqnarray}
The second equation for $\mathcal{H}_{(\lambda-\Delta\lambda)}$ is found
from Eq.~\eqref{G31} by inserting  \eqref{B30}-\eqref{B33}. By
comparing in both equations the coefficients of the operators 
$c^{\dagger}_{{\bf k}m} c_{{\bf k}m}$, 
$
  \left( f_{\mathbf{k}m}^{\dagger} f_{\mathbf{k}m} \right)_{\mathrm{NL}}
$, 
and 
$\left( f_{m}^{\dagger} f_{m} \right)_{\mathrm{L}}$
we find the following relations for the parameters 
at cutoff $\lambda$ and $(\lambda-\Delta\lambda)$
\begin{eqnarray}
  \label{G36}
  \varepsilon_{\mathbf{k},(\lambda-\Delta\lambda)} - 
  \varepsilon_{\mathbf{k},\lambda}
  &=&
  - \, \frac{1}{2} 
  \left[
    \varepsilon_{f,\lambda} + 
    D \left( \Delta_{\mathbf{k},\lambda} - \bar{\Delta}_{\lambda} \right) -
    \varepsilon_{\mathbf{k},\lambda}
  \right]
  \left\{
    \cos \left[ 2 \sqrt{D} A_{\mathbf{k}}(\lambda, \Delta\lambda) \right] - 1
  \right\} \\[1ex]
  && 
  - \,
  \sqrt{D} V_{\mathbf{k}} 
  \sin \left[ 2 \sqrt{D} A_{\mathbf{k}}(\lambda, \Delta\lambda) \right] ,
  \nonumber \\[3ex]
  \label{G37}
  \Delta_{\mathbf{k},(\lambda-\Delta\lambda)} - \Delta_{\mathbf{k},\lambda}
  &=&
  - \, 
  \left[
    \varepsilon_{\mathbf{k},(\lambda-\Delta\lambda)} - 
    \varepsilon_{\mathbf{k},\lambda}
  \right], \\[3ex]
  \label{G38}
  \varepsilon_{f,(\lambda-\Delta\lambda)} - \varepsilon_{f,\lambda}
  &=&
  - \, \frac{1}{D} \frac{1}{N} \sum_{\mathbf{k}}
  \left[
    \varepsilon_{\mathbf{k},(\lambda-\Delta\lambda)} - 
    \varepsilon_{\mathbf{k},\lambda}
  \right]
  \left[
    1 + (\nu_f -1) 
    \left\langle c^{\dagger}_{{\bf k}m} c_{{\bf k}m} \right\rangle
  \right] \\[1ex]
  &&
  +\, \frac{\nu_{f} - 1}{4 D^{3/2}} \, \frac{1}{N}
  \sum_{\mathbf{k}}
  \left\{
    \left[
      \varepsilon_{f,\lambda} + 
      D \left( \Delta_{\mathbf{k},\lambda} - \bar{\Delta}_{\lambda} \right) -
      \varepsilon_{\mathbf{k},\lambda}
    \right]
    \sin\left[2 \sqrt{D} A_{\mathbf{k}}(\lambda, \Delta\lambda) \right]
  \right. \nonumber\\[1ex]
  && \qquad \qquad
  -\, 
  \left.
    2 \sqrt{D} V_{\mathbf{k}} 
    \left\{
      \cos \left[2 \sqrt{D} A_{\mathbf{k}}(\lambda, \Delta\lambda) \right] - 1
    \right\}
  \right\}
  \left\langle
    \hat{f}_{\mathbf{k}m}^{\dagger}c_{\mathbf{k}m}+\mathrm{h.c.}
  \right\rangle \nonumber\\[1ex]
  &&
  - \,
  \frac{\nu_{f} - 1}{2D} \, \frac{1}{N}
  \sum_{\mathbf{k}}
  \left[
    \varepsilon_{f,\lambda} - 
    D \left( \Delta_{\mathbf{k},\lambda} - \bar{\Delta}_{\lambda} \right) -
    \varepsilon_{\mathbf{k},\lambda}
  \right]
  \, A_{\mathbf{k}}(\lambda, \Delta\lambda) \nonumber\\[1ex]
  && \qquad \qquad
  \times \,
  \left\langle
    \hat{f}_{\mathbf{k}m}^{\dagger}c_{\mathbf{k}m}+\mathrm{h.c.}
  \right\rangle, \nonumber \\[3ex]
  \label{G39}
  E_{(\lambda-\Delta\lambda)} - E_{\lambda}
  &=&
  -\, (1-D) \frac{\nu_{f}}{\nu_{f} - 1} N 
  \left[
    \varepsilon_{f,(\lambda-\Delta\lambda)} - \varepsilon_{f,\lambda}
  \right] \\[1ex]
  &&
  -\,
  \frac{1-D}{D} \frac{\nu_{f}}{\nu_{f} - 1} \sum_{\mathbf{k}}
  \left[
    \varepsilon_{\mathbf{k},(\lambda-\Delta\lambda)} - 
    \varepsilon_{\mathbf{k},\lambda}
  \right] . \nonumber
\end{eqnarray}
Note that Eq.~\eqref{G39} for the energy shift $E_{\lambda}$
follows from the comparison of the remaining \textit{c} numbers. 
The renormalization equations \eqref{G36} - \eqref{G38} 
still depend on the expectation values 
$\left\langle c^{\dagger}_{{\bf k}m} c_{{\bf k}m} \right\rangle$, 
$
  \left\langle
    \hat{f}_{\mathbf{k}m}^{\dagger}c_{\mathbf{k}m}+\mathrm{h.c.}
  \right\rangle
$, 
and $D$ [compare \eqref{G23}] which have to be determined first. 
In the following we will discuss an approximate 
evaluation for these quantities which enables us to solve the renormalization 
equations \eqref{G36}-\eqref{G39}. The parameter 
$A_{\mathbf{k}}(\lambda, \Delta\lambda)$ is given by \eqref{G34}. 

A factorization approximation was employed above 
to reduce all renormalization contributions to
operator terms which are bilinear in the fermionic creation and annihilation
operators. In principle, the expectation values are defined 
with the equilibrium distribution of ${\cal H}_{\lambda}$ since 
the renormalization step was done by transforming 
${\cal H}_{\lambda}$ to 
${\cal H}_{(\lambda-\Delta \lambda)}$. 
 ${\cal H}_{\lambda}$ still contains hybridization terms 
between $f$ electrons  and conduction electrons [compare\eqref{G13}]
so that  there is no straight way to evaluate the expectation values.
There are  several approximations possible to circumvent 
this difficulty. Firstly, one can evaluate the expectation values 
by using the unperturbed Hamiltonian ${\mathcal H}_{0,\lambda}$ 
which is diagonal. In this approximation  the 
renormalization equations \eqref{G36}-\eqref{G39} can easily be evaluated
numerically. The result for the renormalized 
$f$ level  found in this way is in good 
agreement with the slave-boson mean field result. 
However,  the quasiparticle energies turn out to be 
discontinuous as function of $\mathbf{k}$. This 
behavior has to be interpreted as an artifact of this simple 
approximation and is caused by the vanishing of expectation values 
$
  \left\langle
    \hat{f}_{\mathbf{k}m}^{\dagger}c_{\mathbf{k}m}+\mathrm{h.c.}
  \right\rangle 
$.
The second possible approximation is more difficult and consists in 
calculating the expectation values with respect to  
the full Hamiltonian ${\mathcal H}$ instead of 
${\mathcal H}_\lambda$. In this case the 
renormalization equations \eqref{G36}-\eqref{G39} can not be evaluated
directly because the expectation values are not known. 
The starting point to find these quantities is the relation
\begin{eqnarray}
  \label{G39a}
  \left\langle {\cal A} \right\rangle &=&
  \frac{
    \mathrm{Tr} 
    \left( {\cal A} e^{-\beta {\cal H}} \right)
  }{
    \mathrm{Tr}\left( e^{-\beta {\cal H}} \right)
  } 
  \,=\, 
  \frac{
    \mathrm{Tr}
    \left( {\cal A}_{\lambda} e^{-\beta {\cal H}_{\lambda}} \right)
  }{
    \mathrm{Tr}\left( e^{-\beta {\cal H}_{\lambda}} \right)}
\end{eqnarray}
which follows from unitarity (for any operator ${\cal A}$). By setting 
up additional renormalization equations for the transformed operators 
${\cal A}_{\lambda}$ one can determine  the expectation values 
$\big< {\cal A} \big>$.  
Note that in the equations for ${\cal A}_\lambda$ 
the unknown expectation values 
enter again so that they have to be solved self-consistently. 
This approach is rather involved but has the advantage that expectation values 
in  \eqref{G36}-\eqref{G39} no longer depend on the cutoff energy 
$\lambda$. 

\bigskip
Renormalization equations for transformed operators also 
have to be used if dynamical correlation functions are evaluated. 
For example, to find
the densities of states of the $f$ electrons
\begin{eqnarray}
  \label{G39b}
  \rho_{f}(\omega) &=& 
  \frac{1}{N} \sum_{{\bf k}m }
  \left\langle\left[
    {\hat f}^{\dagger}_{{\bf k}m}, 
    \delta\left( {\bf L} + \omega \right)
    {\hat f}_{{\bf k}m}
  \right]_{+}\right\rangle
\end{eqnarray}
and of the $c$ electrons
\begin{eqnarray}
  \label{G39c}
  \rho_{c}(\omega) &=& 
  \frac{1}{N} \sum_{{\bf k} m}
  \left\langle\left[
    c^{\dagger}_{{\bf k}m}, 
    \delta\left( {\bf L} + \omega \right)
    c_{{\bf k}m}
  \right]_{+}\right\rangle
\end{eqnarray}
one has to apply the renormalization transformation 
on $\hat{f}^\dagger_{{\bf k}m}$ and
$c^\dagger_{{\bf k}m}$. This will be done in appendix
\ref{one-particle}. Note that in Eqs.~\eqref{G39b} and \eqref{G39c}
the Liouville operator ${\bf L}$ of the full Hamiltonian was introduced
which  is defined by ${\bf L}{\cal A} =[{\cal H}, {\cal A}]$ 
for any operator variable ${\cal A}$.


\section{Analytical solution}
\label{Ana}

Alternatively, one can also find approximate 
analytical solutions for the renormalization 
equations \eqref{G36}-\eqref{G39}. For this purpose three 
approximations have to be used:
\begin{itemize}
  \item[(i)]
  All excitation values are calculated using the full Hamiltonian 
  $\mathcal{H}$ (see the discussion at the end of Sec.~\ref{PAM}), i.e. 
  \begin{eqnarray}
    \label{G40}
    \langle \dots \rangle_{\mathcal{H}_{\lambda}}
    \approx
    \langle \dots \rangle_{\mathcal{H}}
    = \langle \dots \rangle,
  \end{eqnarray}
  so that they are independent from the renormalization parameter $\lambda$.
  \item[(ii)]
  The $\lambda$ dependence of the renormalized $f$ level will be neglected, 
  \begin{eqnarray}
    \label{G41}
    \varepsilon_{f,\lambda} & \approx & \tilde{\varepsilon}_{f}.
  \end{eqnarray}
  The spirit of this approximation is similar to that used in the slave-boson
  theory. There a renormalized $f$ level is used from the very beginning. 
  Within the present treatment one might expect that 
  $\varepsilon_{f,\lambda}$ increases with decreasing $\lambda$ from its 
  initial value $\varepsilon_f$ and reaches its final value 
  $\tilde{\varepsilon}_{f}$ already at finite $\lambda$.
  \item[(iii)]
  The averaged dispersion of $f$ electrons will be neglected,
  \begin{eqnarray}
    \label{G42}
    \bar{\Delta}_{\lambda} &=& 
    \frac{1}{N} \sum_{\mathbf{k}}
    \Delta_{\mathbf{k},\lambda} \,\approx\, 0.
  \end{eqnarray}
\end{itemize}
These approximation enable us to map the renormalization equations of the PAM 
to those of the exactly solvable Fano-Anderson model (see appendix \ref{Fano}).

\subsection{Quasi-particle energies}

Eq.~\eqref{G37} depends on differences of the parameters of the transformed
Hamiltonians at $\lambda$ and $(\lambda - \Delta \lambda)$. Therefore, this 
equation can easily be integrated between a lower cutoff 
$\lambda\rightarrow 0$ and the cutoff $\Lambda$ of the original model. One 
finds 
\begin{eqnarray}
  \label{G43}
  \tilde{\Delta}_{\mathbf{k}} &=&
  - \,
  \left[
    \tilde{\varepsilon}_{\mathbf{k}} - \varepsilon_{\mathbf{k}}
  \right].
\end{eqnarray}
where the initial parameter values \eqref{G15} were used. Furthermore, we have
defined
$
  \tilde{\Delta}_{\mathbf{k}} = \Delta_{\mathbf{k},(\lambda\rightarrow 0)}
$
and
$
  \tilde{\varepsilon}_{\mathbf{k}} = 
  \varepsilon_{\mathbf{k},(\lambda\rightarrow 0)}
$.
Eq.~\eqref{G39} can also be integrated in the same way so that we find 
\begin{eqnarray}
  \label{G44}
  \tilde{E} &=&
  -\, \langle \hat{n}_{i}^{f} \rangle N  
  \left(
    \tilde{\varepsilon}_{f} - \varepsilon_{f}
  \right) - 
  \frac{\langle \hat{n}_{i}^{f} \rangle}{D} \sum_{\mathbf{k}}
  \left(
    \tilde{\varepsilon}_{\mathbf{k}} - \varepsilon_{\mathbf{k}}
  \right).
\end{eqnarray}
Here, again the initial parameter values \eqref{G15} and 
Eq.~\eqref{G23} have been
used. Furthermore, we have defined 
$\tilde{E} = E_{(\lambda\rightarrow 0)}$ and 
$\tilde{\varepsilon}_{f} = \varepsilon_{f,(\lambda\rightarrow 0)}$. The second
term on the r.h.s. of \eqref{G44} vanishes if we use Eq.~\eqref{G43} and
approximation \eqref{G42} so that we obtain
\begin{eqnarray}
  \label{G45}
  \tilde{E} &=&
  -\, \langle \hat{n}_{i}^{f} \rangle N  
  \left(
    \tilde{\varepsilon}_{f} - \varepsilon_{f}
  \right).
\end{eqnarray}

It is more difficult to solve the remaining renormalization  
equations \eqref{G36} and \eqref{G38}. First, the approximations 
\eqref{G40}-\eqref{G42} lead to a decoupling of Eq.~\eqref{G36} for different 
$\mathbf{k}$ values. Thus, by eliminating excitations from large to small 
cutoff values $\lambda$ each $\mathbf{k}$ state is renormalized only once. 
Such a step-like renormalization behavior is also  obtained in the 
Fano-Anderson model (see appendix \ref{Fano}). 
A further similarity to this simpler model is found by inserting the
approximations \eqref{G40}-\eqref{G42} into Eqs.~\eqref{G33} and 
\eqref{G36}
\begin{eqnarray}
  \label{G46}
  \tan \left( 2\sqrt{D} A_{\mathbf{k}} \right)
  &=&
  \frac{
    2 \sqrt{D} V_{\mathbf{k}}
  }{
    \tilde{\varepsilon}_{f} - \varepsilon_{\mathbf{k}}
  }, \\[3ex]
  \label{G47}
  \tilde{\varepsilon}_{\mathbf{k}} - \varepsilon_{\mathbf{k}} &=& 
  \frac{1}{2} 
  \left[
    \cos \left(2 \sqrt{D} A_{\mathbf{k}} \right) - 1
  \right]
  \left[ \varepsilon_{\mathbf{k}} - \tilde{\varepsilon}_{f} \right] - 
  \sqrt{D} V_{\mathbf{k}} \sin\left(2 \sqrt{D} A_{\mathbf{k}} \right).
\end{eqnarray}
Here, the step-like renormalization behavior and the initial parameter values
\eqref{G15} have been used. Note that in the case of a step-like
renormalization the parameters $A_{\mathbf{k}}$ of the generator of the unitary
transformation do not depend on $\lambda$ and $\Delta\lambda$. The above
equations are very similar to those obtained for the Fano-Anderson model 
[compare with Eqs.~\eqref{A11} and \eqref{A10}]. In particular, the
equivalence of the one-particle energies can be seen by replacing 
$\sqrt{D} V_{\mathbf{k}}$ by $V_{\mathbf{k}}$. Moreover, one immediately 
finds from \eqref{G36} and \eqref{G47} the following 
result for the renormalized $c$ electron one-particle energy 
\begin{eqnarray}
  \label{G48}
  \tilde{\varepsilon}_{\mathbf{k}} &=&
  \frac{\tilde{\varepsilon}_{f} +\varepsilon_{\mathbf{k}}}{2} -
  \frac{
    \mathrm{sgn}( \tilde{\varepsilon}_{f} -\varepsilon_{\mathbf{k}} )
  }{2} 
  W_{\mathbf{k}}
\end{eqnarray}
where
\begin{eqnarray}
  \label{G49}
    W_{\mathbf{k}} &=&
  \sqrt{
    \left( \varepsilon_{{\bf k}}-\tilde{\varepsilon}_{f} \right)^{2} + 
    4 D |V_{{\bf k}}|^{2}
  }.
\end{eqnarray}
Obviously, Eqs.~\eqref{G43} and \eqref{G48} also determine the $f$ type
quasi-particle excitation energy which is given by 
\begin{eqnarray}
  \label{G50}
  \tilde{\omega}_{\mathbf{k}} & := &
  \tilde{\varepsilon}_{f} + \tilde{\Delta}_{\mathbf{k}} \, = \,
  \frac{\tilde{\varepsilon}_{f} +\varepsilon_{\mathbf{k}}}{2} +
  \frac{
    \mathrm{sgn}( \tilde{\varepsilon}_{f} -\varepsilon_{\mathbf{k}} )
  }{2} 
  W_{\mathbf{k}}  
\end{eqnarray}
where approximation \eqref{G42} has been used. Thus, we have obtained two 
quasi-particle excitations. According to \eqref{G12} 
and \eqref{G50} the renormalized Hamiltonian reads
\begin{eqnarray}
  \label{G51}
  \tilde{\mathcal{H}} &:=& \mathcal{H}_{(\lambda\rightarrow 0)} \,=\,
  \sum_{\mathbf{k},m}
  \tilde{\omega}_{\mathbf{k}} \,
  f_{\mathbf{k}m}^{\dagger} f_{\mathbf{k}m} 
  + \sum_{{\bf k},m}
  \tilde{\varepsilon}_{{\bf k}} \, c^{\dagger}_{{\bf k}m} c_{{\bf k}m} 
  + \tilde{E} .
\end{eqnarray}
Note that in $\tilde{\cal H}$ the hybridization was completely used up for the 
renormalization of the parameters of ${\cal H}_0$. 
Also the eigenmodes $\hat{f}_{{\bf k}m}$ and $c_{{\bf k}m}$ 
of $\tilde{\cal H}$
do not change their character as function of the 
wave vector due to the presence of the sgn-functions in 
\eqref{G48} and\eqref{G50}.
Instead they remain $f$-like or $c$-like for all values of ${\bf k}$. 
Furthermore, the
one-particle energies \eqref{G48} and \eqref{G50} still depend on two unknown 
quantities, namely, the renormalized \textit{f} level 
$\tilde{\varepsilon}_{f}$ and the expectation value $D$ [see Eq.~\eqref{G23}] 
which have to be determined in the following.

\subsection{Free energy and equations of self-consistency}
First, let us  calculate the averaged \textit{f} electron occupation number 
$\langle \hat{n}_{i}^{f} \rangle$ from the free energy $F$. 
Note that  ${\cal H}_\lambda$ is connected with the original Hamiltonian 
${\cal H}$ by an unitary transformation. Therefore, 
the free energy can also be  evaluated from ${\cal H}_{\lambda}$. 
In particular, the relation  
\begin{eqnarray}
  \label{G52}
  F &=&
  - \,\frac{1}{\beta} \ln {\rm Tr} \,
  e^{-\beta {\cal H}}
  \;=\;
  - \frac{1}{\beta} \ln {\rm Tr} \,
  e^{-\beta \tilde{\mathcal{H}}} =: \tilde{F}
\end{eqnarray}
holds. Because of $\tilde{\mathcal{H}}$ describes an non-interacting Fermi
system the free energy $F$ can be easily calculated
\begin{eqnarray}
  \label{G53}
  F & = & 
  - \frac{\nu_{f}}{\beta}
  \sum_{\mathbf{k}}
  \ln \left[
    1 + e^{ -\beta \tilde{\varepsilon}_{{\bf k}} }
  \right]
  - \frac{\nu_{f}}{\beta}
  \sum_{\mathbf{k}}
  \ln \left[
    1 + e^{ -\beta \tilde{\omega}_{{\bf k}} }
  \right]
  +  \tilde{E}.
\end{eqnarray}
The $f$ electron occupation number is found from $\tilde{F}$ by 
functional derivative
\begin{eqnarray}
  \label{G54}
  \langle \hat{n}_{i}^{f} \rangle
  &=&
  \frac{1}{N}\frac{\partial \tilde{F}}{\partial \varepsilon_{f}} \;=\;
  \frac{1}{N}
  \left\langle
    \frac{\partial \tilde{\mathcal{H}}}{\partial \varepsilon_{f}}
  \right\rangle_{\tilde{\mathcal{H}}} 
\end{eqnarray}
which can be easily performed. We finally obtain a relation of the following 
structure
\begin{eqnarray}
  \label{G55}
  0 &=& 
  \{ \dots \} 
  \left(
    \frac{\partial \tilde{\varepsilon}_{f}}{\partial\varepsilon_{f}}
  \right) + \{ \dots \}
  \left(
    \frac{
      \partial\langle \hat{n}_{i}^{f} \rangle
    }{
      \partial\varepsilon_{f}
    }
  \right).
\end{eqnarray}
We are interested in solutions of the renormalization equations which
describe mixed valence and heavy Fermion behavior. For these cases
the derivatives in Eq.~\eqref{G55} are non-zero so that solutions can be 
found by setting both brace expressions equal to zero. In this way  
the following self-consistent equations for the
renormalized \textit{f} level $\tilde{\varepsilon}_{f}$ and the averaged
\textit{f} electron occupation number $\langle \hat{n}_{i}^{f} \rangle$
are found,
\begin{eqnarray}
  \label{G56}
  \langle \hat{n}_{i}^{f} \rangle &=&
  \frac{\nu_{f}}{N} 
  \sum_{\mathrm{k}}
  f(\tilde{\varepsilon}_{\mathbf{k}})
  \left\{
    \frac{1}{2} + 
    \mathrm{sgn}(\tilde{\varepsilon}_{f} - \varepsilon_{\mathbf{k}})
    \frac{\varepsilon_{\mathbf{k}}-\tilde{\varepsilon}_{f}}{2W_{\mathbf{k}}} \,
  \right\} \\[1ex]
  && +\, 
  \frac{\nu_{f}}{N} \sum_{\mathbf{k}}
  f(\tilde{\omega}_{\mathbf{k}})
  \left\{
    \frac{1}{2} + 
    \mathrm{sgn}(\varepsilon_{\mathbf{k}} - \tilde{\varepsilon}_{f} )
    \frac{\varepsilon_{\mathbf{k}}-\tilde{\varepsilon}_{f}}{2W_{\mathbf{k}}} \,
  \right\}
  \nonumber , \\[3ex]
  \label{G57}
  \tilde{\varepsilon}_{f} - \varepsilon_{f} &=&
  \frac{\nu_{f}-1}{N}
  \sum_{\mathbf{k}}
  \mathrm{sgn}(\tilde{\varepsilon}_{f} - \varepsilon_{\mathbf{k}}) \,
  f(\tilde{\varepsilon}_{\mathbf{k}})
  \frac{|V_{\mathrm{k}}|^{2}}{W_{\mathbf{k}}} \\[1ex]
  && + \, 
  \frac{\nu_{f}-1}{N}
  \sum_{\mathbf{k}}
  \mathrm{sgn}(\varepsilon_{\mathbf{k}} - \tilde{\varepsilon}_{f} ) \,
  f(\tilde{\omega}_{\mathbf{k}})
  \frac{|V_{\mathrm{k}}|^{2}}{W_{\mathbf{k}}} 
  \nonumber.
\end{eqnarray}
Note that these equations are quite similar to those which are 
found in the slave-boson (SB) formalism \cite{Fulde}. In particular, the limit 
$\nu_{f}\rightarrow\infty$ of Eqs.~\eqref{G56} and \eqref{G57} perfectly
agrees with the SB equations.

\subsection{Expectation values}
The remaining expectation values 
$\langle c_{\mathbf{k}m}^{\dagger} c_{\mathbf{k}m}\rangle$ and 
$\langle \hat{f}_{\mathbf{k}m}^{\dagger}c_{\mathbf{k}m} 
+ \mathrm{h.c.}\rangle$
can also be evaluated  from the free energy \eqref{G52} 
\begin{eqnarray}
  \label{G58}
  \left\langle c_{\mathbf{k}m}^{\dagger} c_{\mathbf{k}m} \right\rangle &=&
  \frac{1}{\nu_{f}} 
  \frac{\partial \tilde{F}}{\partial \varepsilon_{\mathbf{k}}} \;=\;
  \frac{1}{\nu_{f}}
  \left\langle
    \frac{\partial \tilde{\mathcal{H}}}{\partial \varepsilon_{\mathbf{k}}}
  \right\rangle_{\tilde{\mathcal{H}}} ,\\[3ex]
  \label{G59}
  \left\langle 
    \hat{f}_{\mathbf{k}m}^{\dagger}c_{\mathbf{k}m} + \mathrm{h.c.}
  \right\rangle
  &=&
  \frac{1}{\nu_{f}} 
  \frac{\partial \tilde{F}}{\partial V_{\mathbf{k}}} \;=\;
  \frac{1}{\nu_{f}}
  \left\langle
    \frac{\partial \tilde{\mathcal{H}}}{\partial V_{\mathbf{k}}}
  \right\rangle_{\tilde{\mathcal{H}}}.
\end{eqnarray}
Both expressions can  be evaluated similarly to \eqref{G54}. 
By using \eqref{G56} and \eqref{G57} we find
\begin{eqnarray}
  \label{G60}
  \left\langle c_{\mathbf{k}m}^{\dagger} c_{\mathbf{k}m}\right\rangle 
  &=&
  \frac{1}{2}
  \left[
    1 - 
    \mathrm{sgn}(\tilde{\varepsilon}_{f} - \varepsilon_{\mathbf{k}}) \, 
    \frac{\varepsilon_{\mathbf{k}}-\tilde{\varepsilon}_{f}}{2W_{\mathbf{k}}}
  \right]
  f(\tilde{\varepsilon}_{\mathbf{k}}) \\[1ex]
  &&
  + \,
  \frac{1}{2}
  \left[
    1 - 
    \mathrm{sgn}(\varepsilon_{\mathbf{k}} - \tilde{\varepsilon}_{f}) \, 
    \frac{\varepsilon_{\mathbf{k}}-\tilde{\varepsilon}_{f}}{2W_{\mathbf{k}}}
  \right]
  f(\tilde{\omega}_{\mathbf{k}}) ,\nonumber\\[3ex]
  \label{G61}
  \left\langle 
    \hat{f}_{\mathbf{k}m}^{\dagger}c_{\mathbf{k}m} + \mathrm{h.c.}
  \right\rangle
  &=&
  - \, 2 \,\mathrm{sgn}(\tilde{\varepsilon}_{f} - \varepsilon_{\mathbf{k}})\,
  \frac{D V_{\mathbf{k}}}{W_{\mathbf{k}}} \,
  f(\tilde{\varepsilon}_{\mathbf{k}}) 
  -
  2 \,\mathrm{sgn}(\varepsilon_{\mathbf{k}} - \tilde{\varepsilon}_{f})\,
  \frac{D V_{\mathbf{k}}}{W_{\mathbf{k}}} \,
  f(\tilde{\omega}_{\mathbf{k}}) .
\end{eqnarray}
Note that also Eqs. \eqref{G60} and \eqref{G61} are very similar 
to the corresponding SB results.

\subsection{One-particle operators and density of states}
\label{density}
Next we calculate the densities of states of the $f$ and $c$
electrons [compare Eqs. \eqref{G39b} and \eqref{G39c}]. For that purpose, we
have to integrate the renormalization equations \eqref{C9} and \eqref{C10} to
determine the transformed one-particle operators. In the case of the
analytical solution, \eqref{C9} and \eqref{C10}  can be exactly
solved if the basic approximations \eqref{G41} and \eqref{G42} are used. 
As already discussed above, in this case the different $\mathbf{k}$ values 
are not coupled with each other and we obtain a step-like
renormalization behavior. Thus, we find
\begin{eqnarray}
  \label{G65}
  \tilde{u}_{\bf k} &=& 
  \cos \left( \sqrt{D} A_{\mathbf{k}} \right)
  \qquad \mbox{and} \qquad
  \tilde{v}_{\bf k} \,=\, 
  \frac{1}{\sqrt{D}} \sin \left( \sqrt{D} A_{\mathbf{k}} \right)
\end{eqnarray}
where the initial parameter values \eqref{C2} were used. Furthermore, we have
defined $\tilde{u}_{\bf k} = u_{{\bf k},(\lambda\rightarrow 0)}$ and 
$\tilde{v}_{\bf k} = v_{{\bf k},(\lambda\rightarrow 0)}$. Combining the
generator of the unitary transformation \eqref{G46}, the normalization
condition \eqref{C3}, and Eq.~\eqref{G65} we finally obtain
\begin{eqnarray}
  \label{G66}
  \left| \tilde{u}_{\mathbf{k}} \right|^{2} &=&
  \frac{1}{2}
  \left\{
    1 - \frac{\varepsilon_{\bf k} - \tilde{\varepsilon}_{f}}{W_{\bf k}}
    \mathrm{sgn}\left( \tilde{\varepsilon}_{f} - \varepsilon_{\bf k} \right)
  \right\}\,\\[1ex]
  \label{G67}
  \left| \tilde{v}_{\mathbf{k}} \right|^{2} &=&
  \frac{1}{2D}
  \left\{
    1 + \frac{\varepsilon_{\bf k} - \tilde{\varepsilon}_{f}}{W_{\bf k}}
    \mathrm{sgn}\left( \tilde{\varepsilon}_{f} - \varepsilon_{\bf k} \right)
  \right\}\,.
\end{eqnarray}
Thus, the coefficients $\tilde{u}_{\mathbf{k}}$ and $\tilde{v}_{\mathbf{k}}$
can be directly calculated from the results of the  
self-consistent equations \eqref{G56} and \eqref{G57}.

To calculate the densities of states \eqref{G39b} and \eqref{G39c} we use the
relation \eqref{G39a} which follows from the unitary of all operators. Thus,
Eqs. \eqref{G39b} and \eqref{G39c} can be rewritten as 
\begin{eqnarray}
  \label{G68}
  \rho_{f}(\omega) &=& 
  \frac{1}{N} \sum_{{\bf k}m}
  \left\langle\left[
    {\hat f}^{\dagger}_{{\bf k}m}(\lambda \rightarrow 0), 
    \delta\left( \tilde{\bf L} + \omega \right)
    {\hat f}_{{\bf k}m}(\lambda \rightarrow 0)
  \right]_{+}\right\rangle_{\tilde{\cal H}}, \\[1ex]
  \label{G69}
  \rho_{c}(\omega) &=& 
  \frac{1}{N} \sum_{{\bf k}m}
  \left\langle\left[
    c^{\dagger}_{{\bf k}m}(\lambda \rightarrow 0), 
    \delta\left( \tilde{\bf L} + \omega \right)
    c_{{\bf k}m}(\lambda \rightarrow 0)
  \right]_{+}\right\rangle_{\tilde{\cal H}}.
\end{eqnarray}
where $\tilde {\bf L}$ is the Liouville operator of the final Hamiltonian 
$\tilde{\cal H}$ which is defined by 
$\tilde{\bf L}{\cal A} = [\tilde{\cal H}, {\cal A}]$ for all operator
variables ${\cal A}$. Due to the structure \eqref{G51} of the final
Hamiltonian $\tilde{\cal H}$, Eqs. \eqref{G68} and \eqref{G69} can be easily
evaluated. Using \eqref{C1}, \eqref{C4}, \eqref{G66}, and \eqref{G67} we
obtain
\begin{eqnarray}
  \label{G70}
  \rho_{f}(\omega) &=& 
  D \frac{1}{N} \sum_{{\bf k}m}
  \left\{
    \left| \tilde{u}_{\bf k} \right|^{2} 
    \delta\left( \omega - \tilde\omega_{\bf k} \right) + 
    D \left| \tilde{v}_{\bf k} \right|^{2} 
    \delta\left( \omega - \tilde\varepsilon_{\bf k} \right)
  \right\},\\[2ex]
  \label{G71}
  \rho_{c}(\omega) &=& 
  \frac{1}{N} \sum_{{\bf k}m}
  \left\{
    \left| \tilde{u}_{\bf k} \right|^{2} 
    \delta\left( \omega - \tilde\varepsilon_{\bf k} \right) + 
    D \left| \tilde{v}_{\bf k} \right|^{2} 
    \delta\left( \omega - \tilde\omega_{\bf k} \right)
  \right\}.
\end{eqnarray}

\begin{figure}
  \begin{center}
    \scalebox{0.8}{
      \includegraphics*[80,420][532,750]{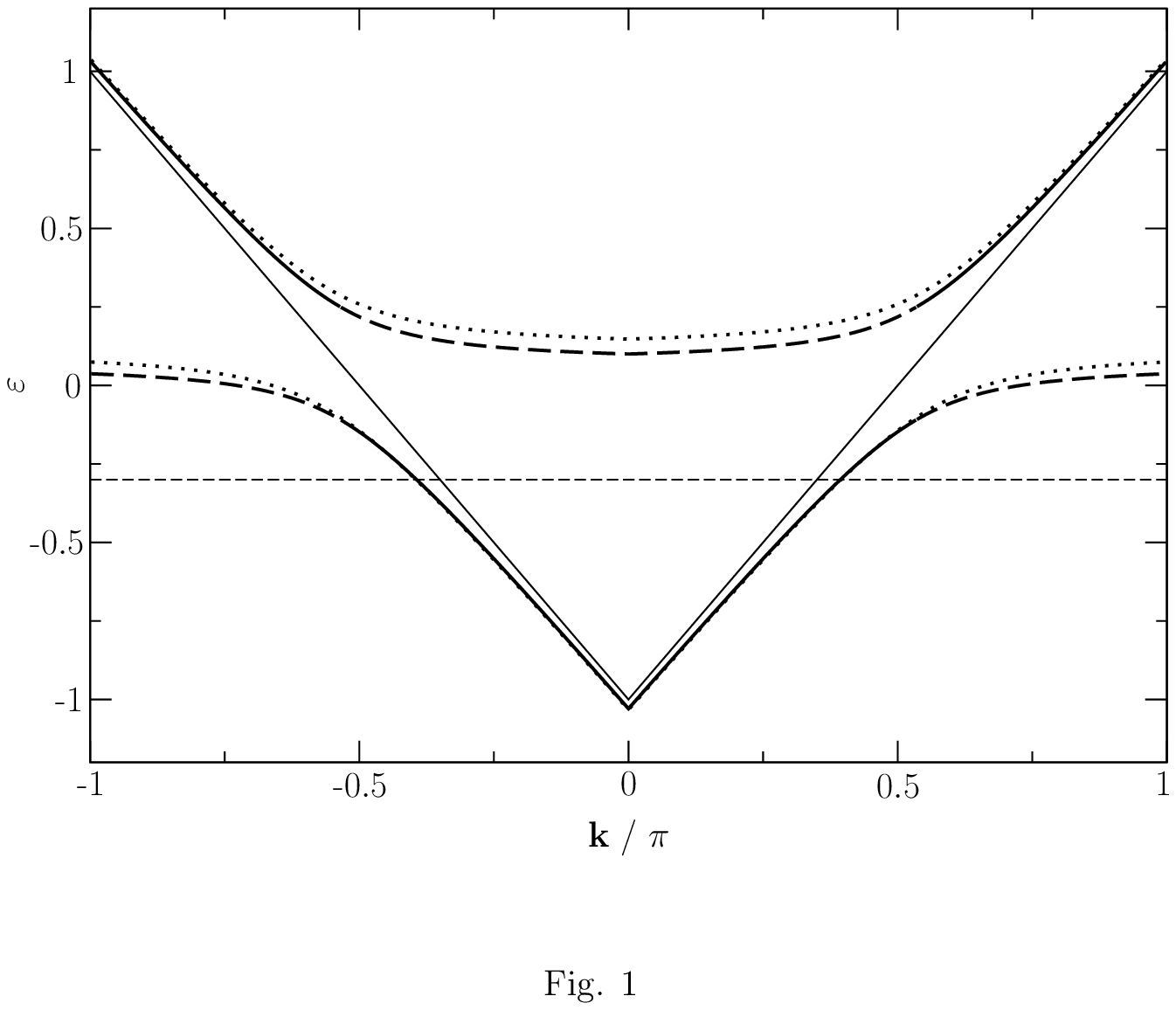}
    }
  \end{center}
  \caption{
    Dispersion relations of an one-dimensional PAM ($N=50000$, $\nu_{f}=4$, 
    $\nu_{f}V^{2}=0.36$, $\varepsilon_{f}=-0.3$, $\mu=0$, T=0.00001). Here, the
    unrenormalized one-particle energies $\varepsilon_{\mathbf{k}}$ and
    $\varepsilon_{f}$ are plotted with full and dashed thin lines. The
    renormalized quasiparticle energies $\tilde{\varepsilon}_{\mathbf{k}}$
    and $\tilde{\omega}_{\mathbf{k}}$ are shown with full and dashed thick 
    lines. Furthermore, the quasiparticle energies of the SB approach
    are drawn by use 
    of dotted lines.
  }
  \label{Fig_1}
\end{figure}

\begin{figure}
  \begin{center}
    \scalebox{0.8}{
      \includegraphics*[70,420][542,750]{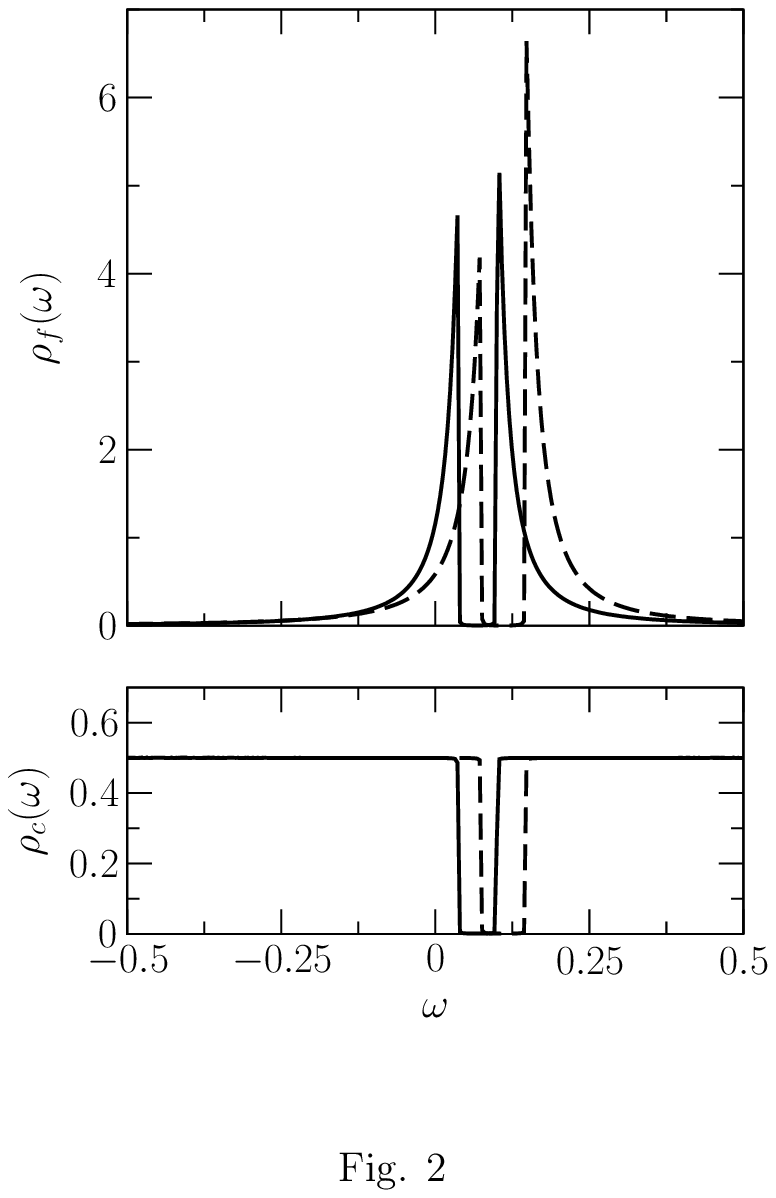}
    }
  \end{center}
  \caption{
    Density of states of the $f$ electrons (upper panel) and of the 
    $c$ electrons (lower panel) where all parameters are chosen as in 
    Fig.~\ref{Fig_1}. A broadening of the $\delta$-functions of $0.0001$
    is used. The results of the analytical PRM solution (SB theory) are drawn 
    as solid (dashed) lines.
  }
  \label{Fig_2}
\end{figure}

\begin{figure}
  \begin{center}
    \scalebox{0.8}{
      \includegraphics*[70,420][542,750]{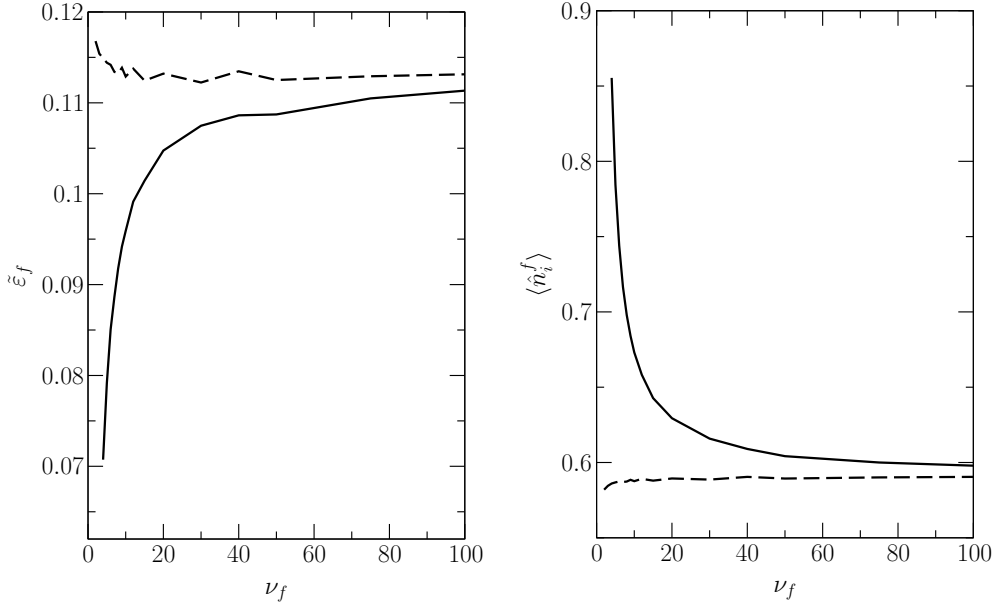}
    }
  \end{center}
  \caption{
    Dependence of the renormalized $f$ level $\tilde{\varepsilon}_{f}$ (left
    panel) and of the averaged $f$ occupation $\langle \hat{n}_{i}^{f}\rangle$
    (right panel) on the degeneracy $\nu_{f}$ of the angular momentum where all
    other parameters are chosen as in Fig.~\ref{Fig_1}. The results of the
    analytical solution (SB theory) are drawn using solid (dashed) lines.
  }
  \label{Fig_3}
\end{figure}

\begin{figure}
  \begin{center}
    \scalebox{0.8}{
      \includegraphics*[70,420][542,750]{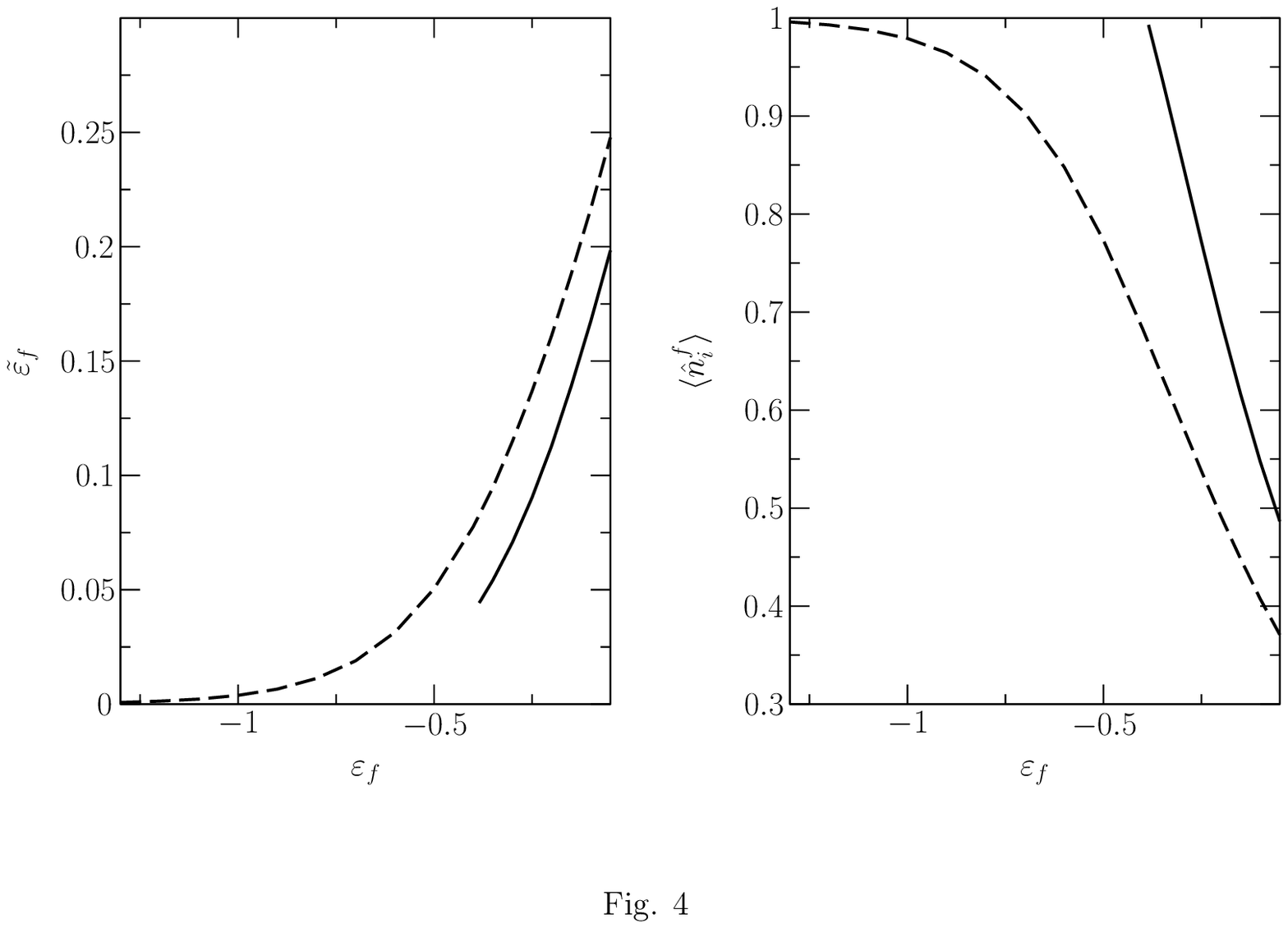}
    }
  \end{center}
  \caption{
    Dependence of the renormalized $f$ level $\tilde{\varepsilon}_{f}$ (left
    panel) and of the averaged $f$ occupation $\langle \hat{n}_{i}^{f}\rangle$
    (right panel) on the original $f$ energy $\varepsilon_{f}$ where all
    other parameters are chosen as in Fig.~\ref{Fig_1}. The results of the
    analytical solution (SB theory) are drawn using solid (dashed) lines.
  }
  \label{Fig_4}
\end{figure}

\begin{figure}
  \begin{center}
    \scalebox{0.8}{
      \includegraphics*[80,400][532,750]{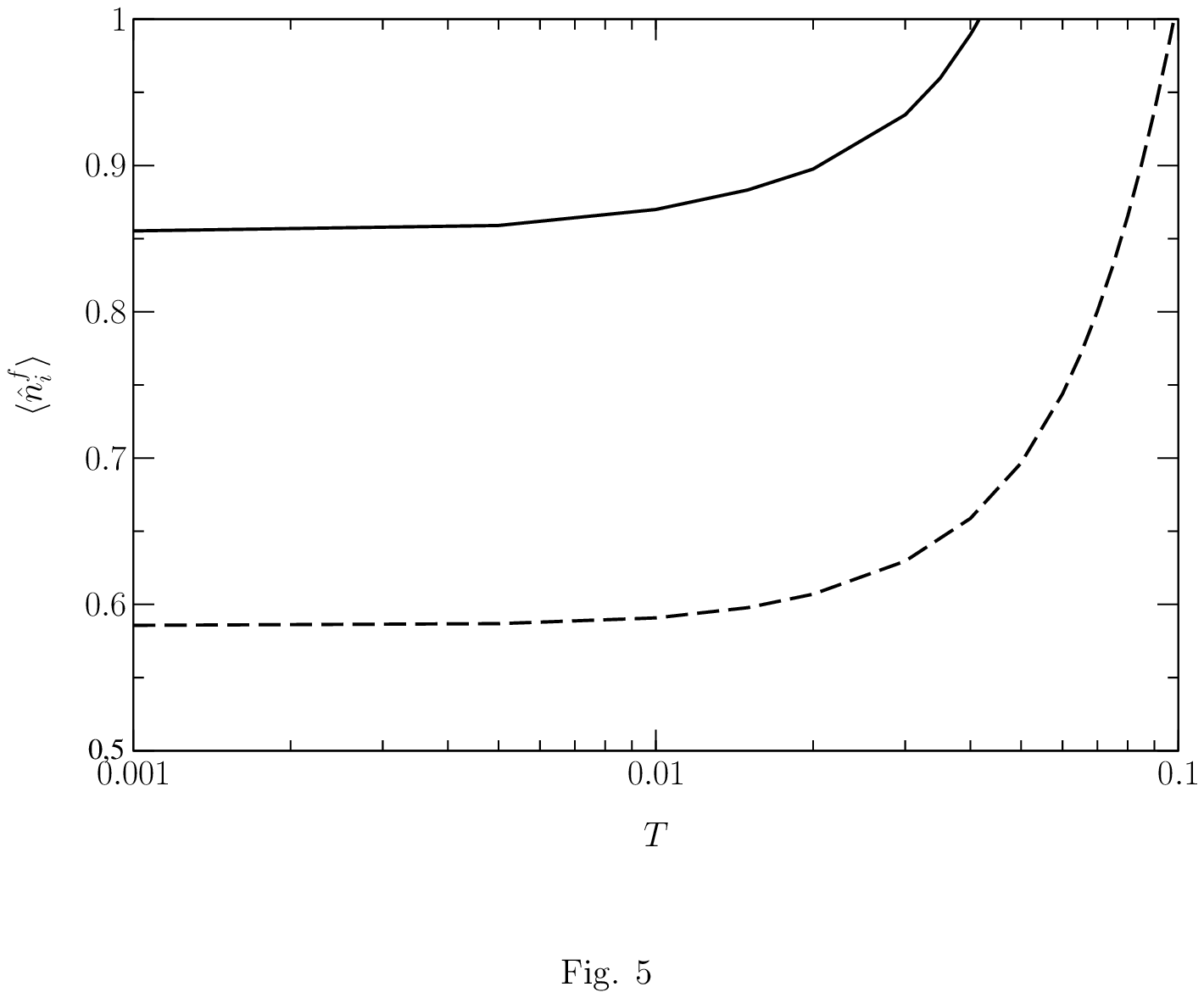}
    }
  \end{center}
  \caption{
    Averaged $f$ occupation $\langle \hat{n}_{i}^{f}\rangle$ as a function of
    temperature $T$ where all other parameters are chosen as in
    Fig.~\ref{Fig_1}. The result of the analytical solution (SB theory) is
    drawn using solid (dashed) line.
  }
  \label{Fig_5}
\end{figure}

\subsection{Results and comparison with slave-boson mean-field theory}
\label{SB}
In this subsection we shall compare the results of our analytical solution
discussed above with those of the slave-boson mean-field (SB) treatment. As
already mentioned, the limit $\nu_{f}\rightarrow\infty$ of the derived
self-consistent equations \eqref{G56} and \eqref{G57} is completely
equivalent to the SB equations. Furthermore, in this limit the expectation
values of our analytical solution [see Eqs.~\eqref{G60} and \eqref{G61}] and
the SB treatment also perfectly agree. Therefore, we want to concentrate on
the case of small degeneracy $\nu_f$. At this point it is important to
notify that we have never exploited an $1/\nu_f$ expansion in the
derivation of the analytical solution of the PAM so 
that it is valid for large as well as small
degeneracy $\nu_f$.

For simplicity, let us consider an one-dimensional PAM with $50000$ lattice
sites, a linear dispersion relation for the conduction electrons, and a 
$\mathbf k$ independent hybridization $V_{\mathbf{k}}=V$ and compare our 
results with those of the 
slave-boson mean-field (SB) theory. In particular, we are interested 
in the dependence of the results on the degeneracy $\nu_{f}$.

At first, let us fix the degeneracy of the angular momentum to $\nu_{f}=4$.
The other parameters are chosen as follows $\nu_{f}V^{2}=0.36$, 
$\varepsilon_{f}=-0.3$, chemical potential $\mu=0$, and $T=0.00001$ where all
energies are given in units of the half bandwidth. As can be seen from
Fig.~\ref{Fig_1}, the renormalized quasi-particle energies, i.e. 
$\tilde{\varepsilon}_{\mathbf{k}}$ and $\tilde{\omega}_{\mathbf{k}}$, 
obtained from 
\eqref{G48}, \eqref{G50}, \eqref{G56}, and \eqref{G57} (full and dashed thick
lines), and the quasi-particle bands of the SB theory (dotted lines) seem to
be quite similar. However, the averaged \textit{f} occupation 
$\langle\hat{n}_{i}^{f}\rangle=0.855$ and the renormalized $f$ level 
$\tilde{\varepsilon}_{f}=0.071$ differ significantly from the SB results 
($\langle\hat{n}_{i}^{f}\rangle=0.586$ and
$\tilde{\varepsilon}_{f}=0.115$). These differences
are mainly caused by the fact that we
have taken into account all $1/\nu_{f}$ corrections which are absent in the 
SB treatment. Note that $1/\nu_{f}$ corrections allow for 
additional renormalization processes which lead 
to a lowering of the free energy for the whole parameter space.

It is well known that the quasi-particles of the SB theory change their 
character as function of ${\mathbf k}$ between a more $f$-like and more
$c$-like behavior. As was mentioned before, 
in the present treatment excitations do not 
change their character as function of ${\mathbf k}$. However, the
quasi-particle energies show jumps in their ${\mathbf k}$ dependence where the
renormalization contributions change their sign from positive to negative
values or  vice versa. Note, that the various parts of the
quasiparticle bands fit perfectly together (see Fig.~\ref{Fig_1}).

As compared to the dispersion relations plotted in Fig.~\ref{Fig_1},
the densities of states of the 
$f$- and $c$-electrons in Fig.~\ref{Fig_2} show much better
the differences between the results 
of the present analytical solution and of the SB treatment. 
In particular,
the smaller value for the renormalized $f$ level 
$\tilde{\varepsilon}_f$ obtained from our PRM treatment leads to much higher 
density of states at the Fermi surface than the SB treatment. Note that such
an enhanced density of states at the Fermi energy is a clear signature of
heavy fermion behavior. 

Next, we discuss the dependence on the degeneracy parameter
$\nu_{f}$. For that purpose we vary $\nu_{f}$ by keeping  
$\nu_{f}V^{2} = 0.36$ fixed.  In contrast to  
the SB results for $\langle\hat{n}_{i}^{f}\rangle$ and 
$\tilde{\varepsilon}_{f}$, which  are almost unchanged 
(see Fig. \ref{Fig_3}), the analytical solutions show a remarkable dependence
on the degeneracy $\nu_{f}$. In particular, 
for small $\nu_{f}$, the $1/\nu_{f}$ corrections included in the PRM approach 
lead to serious deviations from the SB results. From these additional
$1/\nu_{f}$ corrections follows a more pronounced heavy Fermion behavior.
As already mentioned above,
the limit $\nu_{f}\rightarrow\infty$ of our analytical solution perfectly
agrees with the SB theory. To perform this limit one has to replace the
expectation value $D$ by $(1 - \langle\hat{n}_{i}^{f}\rangle)$ so that all
processes are neglected by which a localized electron at an occupied $f$ site
is annihilated and instead a conduction electron is created [compare
discussion below Eq.~\eqref{G32}]. 

In Fig.~\ref{Fig_4}, the renormalized $f$ level $\tilde{\varepsilon}_f$ 
and the 
averaged \textit{f} occupation $\langle\hat{n}_{i}^{f}\rangle$ are plotted as 
functions of the original $f$ energy $\varepsilon_{f}$. The momentum 
degeneracy has been fixed at $\nu_{f}=4$. As is seen, the $1/\nu_{f}$ 
corrections do not only cause a dependence of the results on
the degeneracy $\nu_{f}$ (as shown in Fig.~\ref{Fig_3}) but also lead to a 
reduction of the stability range of heavy Fermion type solutions. 

The $1/\nu_{f}$ corrections also affect the
thermodynamic properties of the system. In Fig.~\ref{Fig_5} the temperature
dependence of the averaged \textit{f} occupation
$\langle\hat{n}_{i}^{f}\rangle$ is shown where $\nu_f$ has been fixed
to $\nu_{f}=4$. We observe that $\langle\hat{n}_{i}^{f}\rangle$ 
goes with increasing temperature much faster to 1 than the
SB results. Thus, the $1/\nu_{f}$ corrections lead to a lowering
of the Kondo temperature
$T_{\mathrm{K}}$ which may be  defined as that temperature at which 
$\langle\hat{n}_{i}^{f}\rangle$ becomes $1$. 

As can be seen from Fig.~\ref{Fig_4}, the analytical PRM 
solution breaks down when the unrenormalized $f$ level becomes smaller 
than some critical values. A similar behavior is also known from the 
SB solution. For instance, the solution for $\big< \hat{n}_i^f \big>$ 
breaks down when for fixed $\varepsilon_f$
the chemical potential $\mu$ is increased beyond  some 
critical value \cite{Franco}.  
The reason for this breakdown is not completely clear.
May be,  it is due to some rough approximation used 
both in the PRM treatment and the SB theory, for instance those  
from Sect.~III.A which have their counterparts in the SB treatment. 
Alternative, the breakdown of the PRM and the SB 
solutions might be a signature of a genuine phase transition. 
Recently it was suggested \cite{Holmes} that in certain systems 
like CeCu$_2$Si$_2$ there might be a transition   
between an intermediate valence regime with 
fluctuating $f$ charges and a regime with integral $f$ charge
when the pressure is decreased. In the integral regime  which 
is described by the Kondo Hamiltonian there is no longer a 
renormalized $f$ level at the Fermi level. This might be the reason that  
the self-consistent solution for $\tilde{\varepsilon}_f$
no longer exists. Note however that such a phase transition  
does not appear in a recent alternative discussion of the PAM 
on the basis of Hubbard operators in Ref.~\onlinecite{Franco}. 
This approach is based on an extended chain approximation and
gives the same quasi-particle energies \eqref{G48} and \eqref{G50}. 
However, it leads to completely different equations for 
the renormalized $\tilde{\varepsilon}_f$ level and for the averaged
\textit{f} occupation $\langle\hat{n}_{i}^{f}\rangle$. Results have been 
found which are very similar to the SB solutions for those 
parameter regimes where the SB solution exists.
In contrast, the PRM solution leads to substantial
deviations from the SB results in particular for small values of
$\nu_f$. Note however that apart from the $1/\nu_f$ corrections 
we have used similar approximations as in the SB theory to derive our
analytical solution.


\section{Conclusions}
\label{Conclusions}

In summary, in this paper we have applied a recently developed
projector-based renormalization method (PRM) to the periodic Anderson model
(PAM) in the limit of infinitely large Coulomb repulsion at $f$
sites. By using an additional  factorization approximation 
we have derived renormalization equations for the parameters of the
Hamiltonian. In this way, the PAM is mapped to a free system consisting 
of two uncorrelated quasiparticle bands. Similar uncorrelated Hamiltonians have
been also derived before by different 
theoretical approaches, such as the Gutzwiller 
projection \cite{Rice} and the SB theory \cite{Coleman,Fulde} where $1/\nu_{f}$
expansions have been exploited. In contrast, the present approach is valid for 
any $\nu_f$. 
Due to the factorization approximation certain expectation values enter
which prevent a direct numerical evaluation of the 
renormalization equations. In principle, the expectation values could be 
determined self-consistently by deriving additional renormalization equations 
also for the operators which enter the expectation values. This has 
not been done in this paper. 

To obtain instead an analytical solution we have used a renormalized 
$f$ level which was assumed to be constant 
during the renormalization process. The spirit of
this approximation is similar to that used in the SB theory
\cite{Coleman,Fulde}. We obtain self-consistent equations for
the renormalized \textit{f} level $\tilde{\varepsilon}_{f}$ and the averaged
\textit{f} occupation $\langle \hat{n}_{i}^{f} \rangle$ which are quite
similar to those of the SB theory \cite{Fulde}. In particular, in the limit
$\nu_f\rightarrow\infty$ our solution perfectly agrees with the SB result
but strongly differs from it for smaller values of $\nu_f$.
To compare our results in more detail 
with those of the SB approach 
we have also considered an one-dimensional PAM with a linear dispersion 
relation of the conduction electrons and a $\mathbf{k}$ independent 
hybridization. Note that the character of the obtained two 
quasi-particle bands of the two treatments
differ. Whereas the quasi-particles of the SB theory change their character as 
function of $\mathbf{k}$ from a more $f$-like to a more $c$-like behavior
and
vice versa the excitations of the PRM treatment do not change their
character. Instead, the quasi-particle energies show jumps 
as function of $\mathbf{k}$ where, however, 
the various parts of the quasi-particle bands
fit perfectly together.
The influence of the degeneracy $\nu_{f}$ has been
studied by varying $\nu_{f}$ with fixed $\nu_{f}V^{2}$. 
Whereas the SB results are almost unchanged, 
our analytical results show a remarkable dependence on the degeneracy $\nu_f$.
Especially, serious deviations are found for small values of $\nu_f$.
   
Finally, from a more technical point of view, note that in 
in this paper the PRM method was applied to
a physical system for the first time  without using any perturbation theory.

\section*{Acknowledgments}

We would like to acknowledge fruitful discussions with K. Meyer,
T. Sommer, and S. Sykora. This work was supported by the 
DFG through the research program  SFB 463, and under Grant No. HU 993/1-1.

\newpage


\begin{appendix}

\section{Renormalization of the exact solvable Fano-Anderson model}
\label{Fano}

In this appendix we illustrate the usefulness of the 
projector-based renormalization method (PRM) for the case of a simple model. 
We apply the approach of Sec.~\ref{PRM} 
to the exactly solvable Fano-Anderson model \cite{Anderson,Fano}. 
This model was already discussed in the framework 
of the present approach in Ref.~\onlinecite{Becker}. However, 
now this will be done in a consequent non-perturbative manner.

\subsection{Model}

The Fano-Anderson model consists of dispersionless $f$ electrons 
which interact with 
conducting electrons. Thereby all correlation effects are neglected.
The Hamiltonian reads
\begin{eqnarray}
  \label{A1}
  {\cal H} &=& {\cal H}_{0} + {\cal H}_{1} , \\[2ex]
  {\cal H}_{0} &=&
  \sum_{{\bf k},m}
  \left(
    \varepsilon_{f} \, f^{\dagger}_{{\bf k}m} f_{{\bf k}m} +
    \varepsilon_{{\bf k}} \, c^{\dagger}_{{\bf k}m} c_{{\bf k}m}
  \right) , \nonumber \\[2ex]
  {\cal H}_{1} &=&
  \sum_{{\bf k},m} V_{{\bf k}}
  \left(
    f_{{\bf k}m}^{\dagger}c_{{\bf k}m} + c_{{\bf k}m}^{\dagger}f_{{\bf k}m}
  \right) . \nonumber
\end{eqnarray}
As in Eq. \eqref{G1} the index $i$ denotes the $f$ sites, ${\bf k}$ is the
wave vector, and $V_{{\bf k}}$ describes the hybridization between conduction 
and localized electrons. The excitation energies $\varepsilon_{{\bf k}}$ and 
$\varepsilon_{f}$ for conduction and localized electrons are measured from the 
chemical potential $\mu$. Both types of electrons 
are assumed to have the same angular momentum index $m$ with values 
$m=1 \dots \nu_{f}$. Of course, the model is easily solved and leads to two 
hybridized bands 
\begin{eqnarray}
  \label{A2}
  {\cal H} &=&
  \sum_{{\bf k},m} \omega_{{\bf k}}^{(\alpha)}
  \alpha_{{\bf k}m}^{\dagger} \alpha_{{\bf k}m} +
  \sum_{{\bf k},m} \omega_{{\bf k}}^{(\beta)}
  \beta_{{\bf k}m}^{\dagger} \beta_{{\bf k}m}
\end{eqnarray}
where
\begin{eqnarray*}
  \omega_{{\bf k}}^{(\alpha,\beta)} &=&
  \frac{\varepsilon_{{\bf k}} + \varepsilon_f}{2}
  \pm \frac{1}{2} W_{\mathbf{k}}, \\
  W_{\mathbf{k}} &=&
  \sqrt{
    \left( \varepsilon_{{\bf k}}-\varepsilon_{f} \right)^{2} + 
    4 |V_{{\bf k}}|^{2}
  } .
\end{eqnarray*}
The eigenmodes $\alpha_{{\bf k}m}^{\dagger}$ and $\beta_{{\bf k}m}^{\dagger}$ 
are given by linear combinations of $c_{{\bf k}m}^{\dagger}$ and 
$f_{{\bf k}m}^{\dagger}$, 
\begin{eqnarray}
  \label{A3}
  \begin{array}{*{7}{c}}
    \alpha_{{\bf k}m}^{\dagger} & = & 
    u_{{\bf k}} \, f_{{\bf k}m}^{\dagger} + v_{{\bf k}} \,
    c_{{\bf k}m}^{\dagger}, &\qquad &
    \beta_{{\bf k}m}^{\dagger} & = &
    -v_{{\bf k}} \, f_{{\bf k}m}^{\dagger} + u_{{\bf k}} \,
    c_{{\bf k}m}^{\dagger} ,  \\[2ex]
    |u_{{\bf k}}|^{2} & = &
    {\displaystyle
      \frac{1}{2}
      \left( 
        1 - \frac{\varepsilon_{{\bf k}}-\varepsilon_{f}}{ W_{\mathbf{k}} }
      \right),
    }
    & \qquad & 
    |v_{{\bf k}}|^{2} & = &
    {\displaystyle
      \frac{1}{2} 
      \left(
        1 + \frac{\varepsilon_{{\bf k}}-\varepsilon_{f}}{ W_{\mathbf{k}} }
      \right) . 
    }
  \end{array}
\end{eqnarray}

\subsection{Renormalization ansatz}

In the renormalization approach we integrate out particle-hole excitations of 
conduction and $f$ electrons which enter due to 
the hybridization term ${\cal H}_{1}$. We expect 
to finally obtain from the PRM an effectively free model. 
The starting point of the method 
is a  renormalized Hamiltonian ${\cal H}_{\lambda}$ which is obtained after
all excitations with energies larger than a given cutoff  $\lambda$ have been 
eliminated. Due to the result of the preceeding section it should have the 
following form
\begin{eqnarray}
  \label{A4}
  {\cal H}_{\lambda} &=& {\cal H}_{0,\lambda} + {\cal H}_{1,\lambda} , \\[2ex]
  {\cal H}_{0, \lambda} &=&
   \sum_{{\bf k},m}
  \left(
    \varepsilon_{{\bf k},\lambda}^{f} \, f^{\dagger}_{{\bf k}m} f_{{\bf k}m} +
    \varepsilon_{{\bf k},\lambda}^{c} \, c^{\dagger}_{{\bf k}m} c_{{\bf k}m}
  \right) ,
  \nonumber \\
 {\cal H}_{1,\lambda}
  &=&
  \sum_{
    \genfrac{}{}{0pt}{1}{
      \genfrac{}{}{0pt}{1}{{\bf k},m}{
        \left|
          \varepsilon_{{\bf k},\lambda}^{c} - \varepsilon_{{\bf k},\lambda}^{f}
        \right|
        \leq \lambda
      }
    }{}
  }
  V_{{\bf k}}
  \left(
    f_{{\bf k}m}^{\dagger}c_{{\bf k}m} + c_{{\bf k}m}^{\dagger}f_{{\bf k}m}
  \right) =  {\bf P}_{\lambda} \, {\cal H}_{1} .
 \nonumber
\end{eqnarray}
As it turns out, no renormalization of the hybridization $ V_{\bf k}$ occurs so
that $V_{\bf k}$ is assumed to be $\lambda$ independent from the beginning.
Like in the exact diagonalization different 
$\mathbf{k}$ states do  not coupled during 
the renormalization process. Thus, by 
eliminating excitations from large to low $\lambda$ values, each 
$\mathbf{k}$ state is renormalized only once leading to  a step like 
renormalization behavior. This  means, for a given cutoff $\lambda$ all 
$\mathbf{k}$ states with excitations 
$|\varepsilon_{\mathbf{k}} - \varepsilon_{f}| > \lambda$ have already been 
renormalized whereas those with 
$|\varepsilon_{\mathbf{k}} -\varepsilon_{f}| < \lambda$ have not. Thus,
${\cal H}_{\lambda}$ can be written as a sum of two parts
${\cal H}_{\lambda} = {\cal H}_{\lambda}^< + {\cal H}_{\lambda}^>$
where
\begin{eqnarray}
  \label{A5}
  {\cal H}_{\lambda}^< &=&
  \sum_{
    \genfrac{}{}{0pt}{1}{
      \genfrac{}{}{0pt}{1}{{\bf k},m}{
        \left|
          \varepsilon_{{\bf k}} - \varepsilon_{f}
        \right|
        \leq \lambda
      }
    }{}
  }
  \left[
    \varepsilon_f \,f_{\mathbf{k}m}^{\dagger}  f_{\mathbf{k}m} +
    \varepsilon_{\mathbf{k}} \, c_{\mathbf{k}m}^{\dagger}  c_{\mathbf{k}m} + 
    V_{\mathbf{k}} \, 
    \left(
      f_{\mathbf{k}m}^{\dagger}  c_{\mathbf{k}m} + 
      c_{\mathbf{k}m}^{\dagger}  f_{\mathbf{k}m}
    \right)
  \right], \\[2ex]
  \label{A6}
  {\cal H}_{\lambda}^> &=&
  \sum_{
    \genfrac{}{}{0pt}{1}{
      \genfrac{}{}{0pt}{1}{{\bf k},m}{
        \left|
          \varepsilon_{{\bf k}} - \varepsilon_{f}
        \right| >  \lambda
      }
    }{}
  }
  \left(
    \tilde{\varepsilon}_{\mathbf{k}}^{f} \, 
    f_{\mathbf{k}m}^{\dagger}  f_{\mathbf{k}m} +
    \tilde{\varepsilon}_{\mathbf{k}}^{c} \, 
    c_{\mathbf{k}m}^{\dagger}  c_{\mathbf{k}m} 
  \right).
\end{eqnarray}
${\cal H}_{\lambda}^<$ is the unchanged part of ${\cal H}_{\lambda}$
whereas  ${\cal H}_{\lambda}^>$ is renormalized due to  the elimination of
excitations $|\varepsilon_{\mathbf{k}} -\varepsilon_{f}|$  larger than 
$\lambda$. $\tilde{\varepsilon}_{\mathbf{k}}^{f}$ and 
$\tilde{\varepsilon}_{\mathbf{k}}^{c}$ denote the renormalized energies.

\subsection{Transformation of the Hamiltonian}

For the explicit evaluation of ${\cal H}_\lambda^>$ let us apply the unitary 
transformation \eqref{G7} on the  original Hamiltonian ${\cal H}$
\begin{eqnarray}
  \label{A7}
  {\cal H}_\lambda &=& e^{X_\lambda} {\cal H} e^{-X_\lambda}.
\end{eqnarray}
For the generator $X_\lambda$ of the unitary transformation 
an exact expression can be given. By inspection of the
perturbation expansion in terms of $V_{\mathbf{k}}$ \cite{Becker} one 
finds that $X_\lambda$ must have the following operator structure
\begin{eqnarray}
  \label{A8}
  X_\lambda &=& 
  \sum_{
    \genfrac{}{}{0pt}{1}{
      \genfrac{}{}{0pt}{1}{{\bf k},m}{
        \left|
          \varepsilon_{{\bf k}} - \varepsilon_{f}
        \right| >  \lambda
      }
    }{}
  }
  A_{\mathbf{k}} 
  \left(
    f_{\mathbf{k}m}^{\dagger} c_{km} - c_{km}^{\dagger} f_{\mathbf{k}m} 
  \right)
\end{eqnarray}
with yet unknown prefactors $A_{\mathbf{k}}$. Eq. \eqref{A8} will be taken as 
ansatz. Then ${\cal H}_{\lambda}^>$ can be easily evaluated 
since only $\mathbf{k}$ values with 
$|\varepsilon_{\mathbf{k}} - \varepsilon_{f}|>\lambda$ renormalize the 
Hamiltonian. Due to the fermionic anticommutator relations  
different $\mathbf{k}$ values are not coupled.
To find  ${\cal H}_{\lambda}^>$, we consider the 
transformation of the various operators. For instance, we obtain  
\begin{eqnarray}
  \label{A9}
  \lefteqn{
    e^{X_\lambda} \, c^{\dagger}_{\mathbf{k}m} c_{\mathbf{k}m} \, 
    e^{-X_\lambda} - c^{\dagger}_{\mathbf{k}m} c_{\mathbf{k}m} \,=\,
  }
  &&\\[2ex]
  &=& 
  \frac{1}{2} \Theta
  \left(
    \left| \varepsilon_{\mathbf{k}} - \varepsilon_f \right| - \lambda
  \right)
  \left\{
    \left[
      \cos\left( 2 A_{\mathbf{k}} \right) - 1
    \right]
    \left(
      c^{\dagger}_{\mathbf{k}m} c_{\mathbf{k}m} - 
      f^{\dagger}_{\mathbf{k}m} f_{\mathbf{k}m}
    \right)
  \right.  \nonumber\\
  && \qquad\qquad\qquad\qquad \qquad +
  \left.
    \sin \left( 2 A_{\mathbf{k}} \right)
    \left(
      f^{\dagger}_{\mathbf{k}m} c_{\mathbf{k}m} + 
      c^{\dagger}_{\mathbf{k}m} f_{\mathbf{k}m}
    \right)
  \right\} \nonumber.
\end{eqnarray}
Similar relations can also be found for the transformations of the
operators
$f^{\dagger}_{\mathbf{k}m} f_{\mathbf{k}m}$ and 
$(
  f^{\dagger}_{\mathbf{k}m} c_{\mathbf{k}m} + 
  c^{\dagger}_{\mathbf{k}m} f_{\mathbf{k}m}
)$. 
Thus, 
${\cal H}_{\lambda}^>$ reads
\begin{eqnarray}
  \label{A10}
  {\cal H}_{\lambda}^> &=&  
  \sum_{
    \genfrac{}{}{0pt}{1}{
      \genfrac{}{}{0pt}{1}{{\bf k},m}{
        \left|
          \varepsilon_{{\bf k}} - \varepsilon_{f}
        \right| >  \lambda
      }
    }{}
  }
  \left\{
    \left[
      \varepsilon_{f} - 
      \frac{1}{2} \left[ \cos (2 A_{\mathbf{k}}) - 1 \right] 
      \left( \varepsilon_{\mathbf{k}} - \varepsilon_{f} \right) + 
      V_{\mathbf{k}} \sin (2 A_{\mathbf{k}})  
    \right]
    f_{\mathbf{k}m}^{\dagger}  f_{\mathbf{k}m} 
  \right.\\
  && \qquad \qquad + 
  \left.
    \left[
      \varepsilon_{\mathbf{k}} + 
      \frac{1}{2} \left[ \cos (2 A_{\mathbf{k}}) - 1 \right] 
      \left( \varepsilon_{\mathbf{k}} - \varepsilon_{f} \right) - 
      V_{\mathbf{k}} \sin (2 A_{\mathbf{k}})  
    \right]
    c_{\mathbf{k}m}^{\dagger}  c_{\mathbf{k}m}
  \right. \nonumber \\[2ex]
  && \qquad\qquad + 
  \left.
    \left[
      V_{\mathbf{k}} + 
      \frac{1}{2} \sin (2 A_{\mathbf{k}}) 
      \left( \varepsilon_{\mathbf{k}} - \varepsilon_{f} \right) + 
      V_{\mathbf{k}} \left[ \cos (2 A_{\mathbf{k}}) - 1 \right] 
    \right]
    \left(
      f_{\mathbf{k}m}^{\dagger}  c_{\mathbf{k}m} + 
      c_{\mathbf{k}m}^{\dagger}  f_{\mathbf{k}m}
    \right)
  \right\}. \nonumber
\end{eqnarray}

In contrast to the expected form \eqref{A6} for ${\cal H}_{\lambda}^{>}$
the expression \eqref{A10} still contains a hybridization  part proportional 
to
$
  (
    f_{\mathbf{k}m}^\dagger c_{\mathbf{k}m} + 
    c_{\mathbf{k}m}^\dagger f_{\mathbf{k}m}
  )
$
with excitation energies larger than $\lambda$. The requirement
\eqref{G8}, ${\bf Q}_{\lambda} {\cal H}_{\lambda} =0$, leads to the following 
condition for $A_{\mathbf{k}}$
 \begin{eqnarray}
  \label{A11}
  \tan(2A_{\mathbf{k}}) &=& 
  \frac{2V_{\mathbf{k}}}{\varepsilon_{f} - \varepsilon_{\mathbf{k}}}.
 \end{eqnarray}
Eq. \eqref{A11} guarantees that the hybridization vanishes in \eqref{A10} and 
${\cal H}_{\lambda}^{>}$ becomes diagonal. Note that according to \eqref{A11}, 
the quantity $A_{\mathbf{k}}$ changes its sign when the 
energy difference  
$\varepsilon_{f} - \varepsilon_{\mathbf{k}}$ changes its sign. By inserting 
\eqref{A11} into \eqref{A10} one finds
\begin{eqnarray}
  \label{A12}
  {\cal H}_{\lambda}^> &=&
  \sum_{
    \genfrac{}{}{0pt}{1}{
      \genfrac{}{}{0pt}{1}{{\bf k},m}{
        \left|
          \varepsilon_{{\bf k}} - \varepsilon_{f}
        \right| >  \lambda
      }
    }{}
  }
  \left(
    \tilde{\varepsilon}_{\mathbf{k}}^{f} \, 
    f_{\mathbf{k}m}^{\dagger}  f_{\mathbf{k}m} +
    \tilde{\varepsilon}_{\mathbf{k}}^{c} \, 
    c_{\mathbf{k}m}^{\dagger}  c_{\mathbf{k}m} 
  \right)
\end{eqnarray}
where the  renormalized energies are given by
\begin{eqnarray}
  \label{A13}
  \tilde{\varepsilon}_{\mathbf{k}}^{f} &=&
  \frac{\varepsilon_{f} +\varepsilon_{\mathbf{k}}}{2} +
  \frac{
    \mathrm{sgn}( \varepsilon_{f} -\varepsilon_{\mathbf{k}} )
  }{2} 
  W_{\mathbf{k}} , \\[2ex]
  \tilde{\varepsilon}_{\mathbf{k}}^{c} &=&
  \frac{\varepsilon_{f} +\varepsilon_{\mathbf{k}}}{2} -
  \frac{
    \mathrm{sgn}( \varepsilon_{f} -\varepsilon_{\mathbf{k}} )
  }{2} 
  W_{\mathbf{k}}. \nonumber
\end{eqnarray}
For $\lambda \rightarrow 0$ the Hamiltonian is completely renormalized. 
The final Hamiltonian 
$ \tilde{{\cal H}}:= {\cal H}_{(\lambda \rightarrow 0)}$ reads
\begin{eqnarray}
  \label{A14}
  \tilde{{\cal H}}&=&
  \sum_{{\bf k},m}
  \left(
    \tilde{\varepsilon}_{\mathbf{k}}^{f} \,  
    f_{\mathbf{k}m}^{\dagger}  f_{\mathbf{k}m} +
    \tilde{\varepsilon}_{\mathbf{k}}^{c}  \, 
    c_{\mathbf{km}}^{\dagger}  c_{\mathbf{k}m} 
  \right).
\end{eqnarray}

Note that the final result \eqref{A14} corresponds to
the diagonal Hamiltonian of eq.\eqref{A2}. In particular, all expectation
values completely agree between the two approaches. (To calculate expectation
values by using Eq.\eqref{A14} one also 
has to transform the operators which enter the
expectation values. For more details see Ref.~\cite{Becker}.) 
However, in contrast to the eigenmodes $\alpha_{\mathbf{k}m}^{\dagger}$ and  
$\beta_{\mathbf{k}m}^{\dagger}$ of \eqref{A2} the present eigenmodes 
$f_{\mathbf{k}m}^\dagger$ and $c_{\mathbf{k}m}^\dagger$ do not change their 
character as function of the wave vector. Remember,  
$\alpha_{\mathbf{k}m}^{\dagger}$ was a more \textit{f}-like excitation for
$\varepsilon_{\mathbf{k}} < \varepsilon_{f}$ and  a more \textit{c}-like
excitation for   $\varepsilon_{\mathbf{k}} > \varepsilon_{f}$, and vice versa 
for $\beta_{\mathbf{k}m}^{\dagger}$. In contrast, the operators
${f}_{\mathbf{k}m}^{\dagger}$ and $c_{\mathbf{k}m}^{\dagger}$ of
\eqref{A14} remain \textit{f}-like and $c$-like for all 
values of ${\bf k}$. In 
return, the $\lambda$ dependent excitation energies 
$\varepsilon_{\mathbf{k},\lambda}^{f}$ and 
$\varepsilon_{\mathbf{k},\lambda}^{c}$ show as function of $\lambda$ a 
step-like behavior at 
$\lambda = \left|\varepsilon_{\mathbf{k}} - \varepsilon_{f}\right|$.
This step-like change guarantees that deviations from the unrenormalized 
energies $\varepsilon_{f}$ and $\varepsilon_{\mathbf{k}}$ 
stay relatively small for all $\mathbf{k}$ values.

\section{Transformation of the operators}
\label{Transformation}

In this appendix we evaluate the transformation from $\lambda$ to 
$(\lambda - \Delta\lambda)$ for the various operator quantities of
Eq.~\eqref{G31}. For example
\begin{eqnarray}
  \label{B1}
  e^{X_{\lambda, \Delta \lambda}}
  c^{\dagger}_{{\bf k}m} c_{{\bf k}m} 
  e^{- X_{\lambda, \Delta \lambda}} 
  &=&
  e^{{\bf X}_{\lambda, \Delta \lambda}} \,
  \left( c^{\dagger}_{{\bf k}m} c_{{\bf k}m} \right) 
  \,=\, 
  \sum_{n=0}^{\infty} 
  \frac{1}{n!}  \, {\bf X}_{\lambda, \Delta \lambda}^{n} \,
  \left( c^{\dagger}_{{\bf k}m} c_{{\bf k}m} \right).
\end{eqnarray}
 Here, a new super-operator ${\bf X}_{\lambda,\Delta \lambda}$ was introduced
which is defined by the commutator of the generator
$X_{\lambda, \Delta \lambda}$ with operators 
${\cal A}$ on which ${\bf X}_{\lambda, \Delta \lambda}$ is applied,
$
  {\bf X}_{\lambda, \Delta \lambda} {\cal A} =
  [X_{\lambda, \Delta \lambda}, {\cal A}]
$.
Furthermore, let us define a new operator $X_{\mathbf{k}m}$ by
\begin{eqnarray*}
  X_{\mathbf{k}m} &=& 
  \hat{f}_{\mathbf{k}m}^{\dagger} c_{\mathbf{k}m} - 
  c_{\mathbf{k}m} \hat{f}_{\mathbf{k}m}^{\dagger}
\end{eqnarray*}
which is an ingredient of the generator of the unitary transformation 
\begin{eqnarray}
  \label{B2}
  X_{\lambda, \Delta \lambda} &=&
  \sum_{\mathbf{k}m} A_{\mathbf{k}}(\lambda, \Delta\lambda) \, 
  \Theta_{\mathbf{k}}(\lambda,\Delta\lambda) \, 
  X_{\mathbf{k}m}.
\end{eqnarray}

We have to evaluate various commutators
\begin{eqnarray}
  \label{B3}
  \left[
    X_{\mathbf{k}'m'}, c_{\mathbf{k}m}^{\dagger} c_{\mathbf{k}}
  \right]
  & = &
  \delta_{\mathbf{k}', \mathbf{k}} \, \delta_{m',m}
  \left(
    \hat{f}_{\mathbf{k}m}^{\dagger} c_{\mathbf{k}m} + \mathrm{h.c.}
  \right), \\[3ex]
  \label{B4}
  \left[
    X_{\mathbf{k}'m'}, 
    \left( \hat{f}_{m}^{\dagger} \hat{f}_{m} \right)_{\mathrm{L}}
  \right] 
  & = &
  - \frac{\delta_{m',m}}{N} 
  \left(
    \hat{f}_{\mathbf{k}'m}^{\dagger} c_{\mathbf{k}'m} + \mathrm{h.c.}
  \right),\\[3ex]
  \label{B5}
  \left[
    X_{\mathbf{k}'m'}, 
    \left( f_{m}^{\dagger} f_{m} \right)_{\mathrm{L}}
  \right] 
  & = &
  \left[
    X_{\mathbf{k}'m'}, 
    \left( \hat{f}_{m}^{\dagger} \hat{f}_{m} \right)_{\mathrm{L}}
  \right] , \\[3ex]
  \label{B6}
  \left[
    X_{\mathbf{k}'m'}, 
    \left( 
      \hat{f}_{\mathbf{k}m}^{\dagger} \hat{f}_{\mathbf{k}m} 
    \right)_{\mathrm{NL}} 
  \right] 
  & = &
  - \frac{\delta_{m',m}}{N^{3/2}} \sum_{i,j(\neq i)}
  \left\{
    e^{i\mathbf{k} \mathbf{R}_{i}} \,
    e^{i (\mathbf{k}'-\mathbf{k}) \mathbf{R}_{j}} \, 
    \hat{f}_{im}^{\dagger} {\cal D}_{jm} c_{\mathbf{k}'m} + 
    \mathrm{h.c.}
  \right\}, \\[3ex]
  \label{B7}
  \left[
    X_{\mathbf{k}'m'}, 
    \left( 
      f_{\mathbf{k}m}^{\dagger} f_{\mathbf{k}m} 
    \right)_{\mathrm{NL}} 
  \right] 
  & = &
  - \frac{\delta_{m',m}}{N^{3/2}} \sum_{i,j(\neq i)}
  \left\{
    e^{i\mathbf{k} \mathbf{R}_{i}} \,
    e^{i (\mathbf{k}'-\mathbf{k}) \mathbf{R}_{j}} \, 
    f_{im}^{\dagger} {\cal D}_{jm} c_{\mathbf{k}'m} + 
    \mathrm{h.c.}
  \right\}, \\[3ex]
  \label{B8}
  \left[
    X_{\mathbf{k}'m'}, 
    \left( 
      \hat{f}_{\mathbf{k}m}^{\dagger} f_{\mathbf{k}m} + \mathrm{h.c.}
    \right)_{\mathrm{NL}} 
  \right]
  & = &
  \left[
    X_{\mathbf{k}'m'}, 
    \left( 
      \hat{f}_{\mathbf{k}m}^{\dagger} \hat{f}_{\mathbf{k}m} 
    \right)_{\mathrm{NL}} 
  \right] 
  +
  \left[
    X_{\mathbf{k}'m'}, 
    \left( 
      f_{\mathbf{k}m}^{\dagger} f_{\mathbf{k}m} 
    \right)_{\mathrm{NL}} 
  \right],
\end{eqnarray}
\begin{eqnarray}
  \label{B9}
  \left[
    X_{\mathbf{k}'m'}, 
    \left(
      \hat{f}_{\mathbf{k}m}^{\dagger} c_{\mathbf{k}m} + \mathrm{h.c.}
    \right)
  \right]
  &=&
  \delta_{m',m}
  \left\{
    2\delta_{\mathbf{k}',\mathbf{k}} 
    \hat{f}_{\mathbf{k}m}^{\dagger} \hat{f}_{\mathbf{k}m} - 
    \left[
      c_{\mathbf{k}m}^{\dagger} c_{\mathbf{k}'m} 
      {\cal D}_{m}(\mathbf{k}'-\mathbf{k}) + \mathrm{h.c.}
    \right] 
  \right\}, \\[3ex]
  \left[
    X_{\mathbf{k}'m'}, 
    \left( f_{\mathbf{k}m}^{\dagger} c_{\mathbf{k}m} + \mathrm{h.c.} \right)
  \right]
  &=&
  \delta_{m',m}
  \left\{
    2\delta_{\mathbf{k}',\mathbf{k}} 
    \left(
      \hat{f}_{\mathbf{k}m}^{\dagger} f_{\mathbf{k}m} + \mathrm{h.c.}
    \right) -
    \left[
      c_{\mathbf{k}m}^{\dagger} c_{\mathbf{k}'m} 
      {\cal D}_{m}(\mathbf{k}'-\mathbf{k})  + \mathrm{h.c.}
    \right] 
  \right\}
  \nonumber\\
  &&  \label{B10}
\end{eqnarray}
where we neglect all spin-flip contributions. In Eqs.~\eqref{B9} and
\eqref{B10} Fourier transformed quantities are introduced
\begin{eqnarray}
  \label{B11}
  {\cal D}_{m}(\mathbf{k}) &=& 
  \frac{1}{N} \sum_{j} e^{i\mathbf{k} \mathbf{R}_{j}} \, {\cal D}_{jm}.
\end{eqnarray}
Furthermore, we have defined
\begin{eqnarray}
  \label{B11a}
  \left( \hat{f}^{\dagger}_{m} \hat{f}_{m} \right)_{\mathrm{L}} &:=&
  \frac{1}{N} \sum_{\mathbf{k}} 
  \hat{f}_{\mathbf{k}m}^{\dagger} \hat{f}_{\mathbf{k}m} \,=\,
  \frac{1}{N} \sum_{i} \hat{f}_{im} ^{\dagger} \hat{f}_{im}, \\
  \label{B11b}
  \left(
    \hat{f}_{\mathbf{k}m}^{\dagger} \hat{f}_{\mathbf{k}m} 
  \right)_{\mathrm{NL}}  
  &:=&
  \frac{1}{N} \sum_{i,j(\neq i)} \hat{f}^{\dagger}_{im} \hat{f}_{jm} \,
  e^{i{\bf k}({\bf R}_i -{\bf R}_j)}
  \,=\,
  \hat{f}_{\mathbf{k}m}^{\dagger} \hat{f}_{\mathbf{k}m} -
  \left( \hat{f}^{\dagger}_{m} \hat{f}_{m} \right)_{\mathrm{L}}, \\
  \label{B11c}
  \left(
    \hat{f}_{\mathbf{k}m}^{\dagger} f_{\mathbf{k}m} + \mathrm{h.c.}
  \right)_{\mathrm{NL}}
  &:=&
  \frac{1}{N} \sum_{i,j(\neq i)} 
  \left[
    \hat{f}^{\dagger}_{im} f_{jm} \,
    e^{i{\bf k}({\bf R}_i -{\bf R}_j)} + \mathrm{h.c.}
  \right].
\end{eqnarray}

We are interested in contributions which renormalize the 
parameters of the Hamiltonian ${\cal H}_\lambda$ according to Eqs. \eqref{G12} 
and \eqref{G13}. Therefore,  an additional factorization has to be carried out 
in Eqs.~\eqref{B6}, \eqref{B7}, \eqref{B9}, and \eqref{B10} and to keep only
operator terms which appear also in  ${\cal H}_\lambda$. By neglecting more
complex operators, namely spin-flip terms,  Eqs.~\eqref{B6}, \eqref{B7}, 
\eqref{B9}, and \eqref{B10} can be replaced by
\begin{eqnarray}
  \label{B12}
  \left[
    X_{\mathbf{k}'m'}, 
    \left( 
      \hat{f}_{\mathbf{k}m}^{\dagger} \hat{f}_{\mathbf{k}m} 
    \right)_{\mathrm{NL}} 
  \right] 
  & = &
  - \delta_{m,m'}
  \left( \delta_{\mathbf{k},\mathbf{k'}} - \frac{1}{N} \right)
  \left\{
    D \left(
      \hat{f}_{\mathbf{k'}m}^{\dagger} c_{\mathbf{k'}m} + \mathrm{h.c.}
    \right)
  \phantom{\sum_{\tilde{m}(\not=m)} }
  \right. \\[1ex]
  && 
  \left.
    \quad + 
    \left\langle
      \hat{f}_{\mathbf{k'}m}^{\dagger} c_{\mathbf{k'}m} + \mathrm{h.c.}
    \right\rangle
    \left[
      1 - D - 
      \sum_{\tilde{m}(\not=m)} 
      \left( 
        \hat{f}_{\tilde{m}}^{\dagger} \hat{f}_{\tilde{m}} 
      \right)_{\mathrm{L}}
    \right]
  \right\}  \nonumber, \\[3ex]
  \label{B13}
  \left[
    X_{\mathbf{k}'m'}, 
    \left( 
      f_{\mathbf{k}m}^{\dagger} f_{\mathbf{k}m} 
    \right)_{\mathrm{NL}} 
  \right] 
  & = &
  - \delta_{m,m'}
  \left( \delta_{\mathbf{k},\mathbf{k'}} - \frac{1}{N} \right)
  \left\{
    D \left(
      f_{\mathbf{k'}m}^{\dagger} c_{\mathbf{k'}m} + \mathrm{h.c.}
    \right)
  \phantom{\sum_{\tilde{m}(\not=m)} }
  \right. \\[1ex]
  && 
  \left.
    \quad + 
    \left\langle
      \hat{f}_{\mathbf{k'}m}^{\dagger} c_{\mathbf{k'}m} + \mathrm{h.c.}
    \right\rangle
    \left[
      1 - D - 
      \sum_{\tilde{m}(\not=m)} 
      \left(
        \hat{f}_{\tilde{m}}^{\dagger} \hat{f}_{\tilde{m}}
      \right)_{\mathrm{L}}
    \right]
  \right\}  \nonumber,\\[3ex]
  \label{B14}
  \left[
    X_{\mathbf{k}'m'}, 
    \left(
      \hat{f}_{\mathbf{k}m}^{\dagger} c_{\mathbf{k}m} + \mathrm{h.c.}
    \right)
  \right]
  &=&
  2 \,\delta_{m,m'} \delta_{\mathbf{k},\mathbf{k'}}
  \left\{
    \left(
      \hat{f}_{\mathbf{k}m}^{\dagger} \hat{f}_{\mathbf{k}m}
    \right)_{\mathrm{NL}} + 
    \left( \hat{f}_{m}^{\dagger} \hat{f}_{m} \right)_{\mathrm{L}} - 
    D c_{\mathbf{k}m}^{\dagger} c_{\mathbf{k}m} 
    \phantom{\sum_{\tilde{m}(\not=m)}}
  \right. \\[1ex]
  &&
  \quad - 
  \left.
    \left\langle c_{\mathbf{k}m}^{\dagger} c_{\mathbf{k}m} \right\rangle
    \left[
      1 - D - 
      \sum_{\tilde{m}(\not=m)} 
      \left(
        \hat{f}_{\tilde{m}}^{\dagger} \hat{f}_{\tilde{m}}
      \right)_{\mathrm{L}}
    \right]
  \right\} \nonumber,
\end{eqnarray}
\begin{eqnarray}
  \label{B15}
  \left[
    X_{\mathbf{k}'m'}, 
    \left(
      f_{\mathbf{k}m}^{\dagger} c_{\mathbf{k}m} + \mathrm{h.c.}
    \right)
  \right]
  &=&
  2 \,\delta_{m,m'} \delta_{\mathbf{k},\mathbf{k'}}
  \left\{
    \frac{1}{2}\left(
      \hat{f}_{m}^{\dagger} f_{m} + \mathrm{h.c.}
    \right)_{\mathrm{NL}} + 
    \left( \hat{f}_{m}^{\dagger} \hat{f}_{m} \right)_{\mathrm{L}}
    \phantom{\sum_{\tilde{m}(\not=m)}}
  \right. \\[1ex]
  &&
  \quad - 
  \left.
    D c_{\mathbf{k}m}^{\dagger} c_{\mathbf{k}m} - 
    \left\langle c_{\mathbf{k}m}^{\dagger} c_{\mathbf{k}m} \right\rangle
    \left[
      1 - D - 
      \sum_{\tilde{m}(\not=m)} 
      \left(
        \hat{f}_{\tilde{m}}^{\dagger} \hat{f}_{\tilde{m}}
      \right)_{\mathrm{L}}
    \right]
  \right\} \nonumber
\end{eqnarray}
where 
$
  \left\langle
    f_{\mathbf{k}m}^{\dagger} c_{\mathbf{k}m} + \mathrm{h.c.}
  \right\rangle
  \approx
  \left\langle
    \hat{f}_{\mathbf{k}m}^{\dagger} c_{\mathbf{k}m} + \mathrm{h.c.}
  \right\rangle
$
has been used. Due to this factorization, certain expectation values enter 
Eqs.~\eqref{B12}-\eqref{B10} which have to be evaluated separately (compare 
the discussion in subsection \ref{ren_equation}). By using 
Eqs.~\eqref{B3}-\eqref{B5} and \eqref{B12}-\eqref{B15} one finds from 
\eqref{B2} for the corresponding commutators formed  with the generator 
$X_{\lambda,\Delta\lambda}$ 
\begin{eqnarray}
  \label{B16}
  \left[
    X_{\lambda,\Delta\lambda}, c_{\mathbf{k}m}^{\dagger} c_{\mathbf{k}}
  \right]
  & = &
  \Theta_{\mathbf{k}}(\lambda,\Delta\lambda) \, 
  A_{\mathbf{k}}(\lambda, \Delta\lambda) \,
  \left(
    \hat{f}_{\mathbf{k}m}^{\dagger} c_{\mathbf{k}m} + \mathrm{h.c.}
  \right), \\[3ex]
  \label{B17}
  \left[
    X_{\lambda,\Delta\lambda}, 
    \left( \hat{f}_{m}^{\dagger} \hat{f}_{m} \right)_{\mathrm{L}}
  \right] 
  & = &
  - \,
  \frac{1}{N} \sum_{\mathbf{k}'}
  \Theta_{\mathbf{k}'}(\lambda,\Delta\lambda) \, 
  A_{\mathbf{k}}(\lambda, \Delta\lambda) \,
  \left(
    \hat{f}_{\mathbf{k}'m}^{\dagger} c_{\mathbf{k}'m} + \mathrm{h.c.}
  \right), \\[3ex]
  \label{B18}
  \left[
    X_{\lambda,\Delta\lambda}, 
    \left( f_{m}^{\dagger} f_{m} \right)_{\mathrm{L}}
  \right] 
  & = &
  \left[
    X_{\lambda,\Delta\lambda}, 
    \left( \hat{f}_{m}^{\dagger} \hat{f}_{m} \right)_{\mathrm{L}}
  \right],
\end{eqnarray}
\begin{eqnarray}
  \label{B19}
  \left[
    X_{\lambda,\Delta\lambda}, 
    \left( 
      \hat{f}_{\mathbf{k}m}^{\dagger} \hat{f}_{\mathbf{k}m} 
    \right)_{\mathrm{NL}} 
  \right]
  & = &
  - \, \Theta_{\mathbf{k}}(\lambda,\Delta\lambda) \, 
  A_{\mathbf{k}}(\lambda, \Delta\lambda)
  \left\{
    D
    \left(
      \hat{f}_{\mathbf{k}m}^{\dagger} c_{\mathbf{k}m} + \mathrm{h.c.}
    \right) 
    \phantom{\sum_{\tilde{m}(\not=m)}}
  \right.\\[1ex]
  &&
  \left.
    \qquad\quad +  
    \left\langle
      \hat{f}_{\mathbf{k}m}^{\dagger} c_{\mathbf{k}m} + \mathrm{h.c.}
    \right\rangle
    \left[
      1 - D - 
      \sum_{\tilde{m}(\not=m)} 
      \left(
        \hat{f}_{\tilde{m}}^{\dagger} \hat{f}_{\tilde{m}}
      \right)_{\mathrm{L}}
    \right]
  \right\}  \nonumber \\[1ex]
  &&
  + \frac{1}{N} \sum_{\mathbf{k'}}
  \Theta_{\mathbf{k'}}(\lambda,\Delta\lambda) \, 
  A_{\mathbf{k'}}(\lambda, \Delta\lambda)
  \left\{
    D
    \left(
      \hat{f}_{\mathbf{k'}m}^{\dagger} c_{\mathbf{k'}m} + \mathrm{h.c.}
    \right) 
    \phantom{\sum_{\tilde{m}(\not=m)}}
  \right.\nonumber\\[1ex]
  &&
  \left.
    \qquad\quad +  
    \left\langle
      \hat{f}_{\mathbf{k'}m}^{\dagger} c_{\mathbf{k'}m} + \mathrm{h.c.}
    \right\rangle
    \left[
      1 - D - 
      \sum_{\tilde{m}(\not=m)} 
      \left(
        \hat{f}_{\tilde{m}}^{\dagger} \hat{f}_{\tilde{m}}
      \right)_{\mathrm{L}}
    \right]
  \right\}  \nonumber,
\end{eqnarray}
\begin{eqnarray}
  \label{B20}
  \left[
    X_{\lambda,\Delta\lambda}, 
    \left( 
      f_{\mathbf{k}m}^{\dagger} f_{\mathbf{k}m} 
    \right)_{\mathrm{NL}} 
  \right]
  & = &
  - \, \Theta_{\mathbf{k}}(\lambda,\Delta\lambda) \, 
  A_{\mathbf{k}}(\lambda, \Delta\lambda)
  \left\{
    D
    \left(
      f_{\mathbf{k}m}^{\dagger} c_{\mathbf{k}m} + \mathrm{h.c.}
    \right) 
    \phantom{\sum_{\tilde{m}(\not=m)}}
  \right.\\[1ex]
  &&
  \left.
    \qquad\quad +  
    \left\langle
      \hat{f}_{\mathbf{k}m}^{\dagger} c_{\mathbf{k}m} + \mathrm{h.c.}
    \right\rangle
    \left[
      1 - D - 
      \sum_{\tilde{m}(\not=m)} 
      \left(
        \hat{f}_{\tilde{m}}^{\dagger} \hat{f}_{\tilde{m}}
      \right)_{\mathrm{L}}
    \right]
  \right\}  \nonumber \\[1ex]
  &&
  + \frac{1}{N} \sum_{\mathbf{k'}}
  \Theta_{\mathbf{k'}}(\lambda,\Delta\lambda) \, 
  A_{\mathbf{k'}}(\lambda, \Delta\lambda)
  \left\{
    D
    \left(
      f_{\mathbf{k'}m}^{\dagger} c_{\mathbf{k'}m} + \mathrm{h.c.}
    \right) 
    \phantom{\sum_{\tilde{m}(\not=m)}}
  \right.\nonumber\\[1ex]
  &&
  \left.
    \qquad\quad +  
    \left\langle
      \hat{f}_{\mathbf{k'}m}^{\dagger} c_{\mathbf{k'}m} + \mathrm{h.c.}
    \right\rangle
    \left[
      1 - D - 
      \sum_{\tilde{m}(\not=m)} 
      \left(
        \hat{f}_{\tilde{m}}^{\dagger} \hat{f}_{\tilde{m}}
      \right)_{\mathrm{L}}
    \right]
  \right\}  \nonumber,
\end{eqnarray}
\begin{eqnarray}
  \label{B21}
  \left[
    X_{\lambda,\Delta\lambda}, 
    \left( 
      \hat{f}_{\mathbf{k}m}^{\dagger} f_{\mathbf{k}m} + \mathrm{h.c.}
    \right)_{\mathrm{NL}} 
  \right]
  &=&
  \left[
    X_{\lambda,\Delta\lambda}, 
    \left( 
      \hat{f}_{\mathbf{k}m}^{\dagger} \hat{f}_{\mathbf{k}m} 
    \right)_{\mathrm{NL}} 
  \right]
  +
  \left[
    X_{\lambda,\Delta\lambda}, 
    \left( 
      f_{\mathbf{k}m}^{\dagger} f_{\mathbf{k}m} 
    \right)_{\mathrm{NL}} 
  \right]
\end{eqnarray}
\begin{eqnarray}
  \label{B22}
  \left[
    X_{\lambda,\Delta\lambda}, 
    \hat{f}_{\mathbf{k}m}^{\dagger} c_{\mathbf{k}m} + \mathrm{h.c.}
  \right]
  & = &
  2 \, \Theta_{\mathbf{k}}(\lambda,\Delta\lambda) \, 
  A_{\mathbf{k}}(\lambda, \Delta\lambda)
  \left\{
    \left(
      \hat{f}_{\mathbf{k}m}^{\dagger} \hat{f}_{\mathbf{k}m}
    \right)_{\mathrm{NL}} +
    \left( \hat{f}_{m}^{\dagger} \hat{f}_{m} \right)_{\mathrm{L}}
    \phantom{\sum_{\tilde{m}(\not=m)}}
  \right.\\[1ex]
  &&
  \left.
    \qquad\quad -
    D c_{\mathbf{k}m}^{\dagger} c_{\mathbf{k}m} -
    \left\langle c_{\mathbf{k}m}^{\dagger} c_{\mathbf{k}m} \right\rangle -
    \left[
      1 - D - 
      \sum_{\tilde{m}(\not=m)} 
      \left(
        \hat{f}_{\tilde{m}}^{\dagger} \hat{f}_{\tilde{m}}
      \right)_{\mathrm{L}}
    \right]
  \right\}, \nonumber\\[3ex]
  \label{B23}
  \left[
    X_{\lambda,\Delta\lambda}, 
    f_{\mathbf{k}m}^{\dagger} c_{\mathbf{k}m} + \mathrm{h.c.}
  \right]
  & = &
  2 \, \Theta_{\mathbf{k}}(\lambda,\Delta\lambda) \, 
  A_{\mathbf{k}}(\lambda, \Delta\lambda)
  \left\{
    \frac{1}{2}\left(
      \hat{f}_{\mathbf{k}m}^{\dagger} f_{\mathbf{k}m} + \mathrm{h.c.}
    \right)_{\mathrm{NL}} +
    \left( \hat{f}_{m}^{\dagger} \hat{f}_{m} \right)_{\mathrm{L}}
    \phantom{\sum_{\tilde{m}(\not=m)}}
  \right.\\[1ex]
  &&
  \left.
    \qquad\quad -
    D c_{\mathbf{k}m}^{\dagger} c_{\mathbf{k}m} -
    \left\langle c_{\mathbf{k}m}^{\dagger} c_{\mathbf{k}m} \right\rangle -
    \left[
      1 - D - 
      \sum_{\tilde{m}(\not=m)} 
      \left(
        \hat{f}_{\tilde{m}}^{\dagger} \hat{f}_{\tilde{m}}
      \right)_{\mathrm{L}}
    \right]
  \right\}. \nonumber
\end{eqnarray}
Therefore, all operators terms appearing on the r.h.~sides of
Eqs.~\eqref{B16}-\eqref{B23} are traced back to a bilinear form. 
This property will enable us to evaluate higher order commutators with 
$X_{\lambda, \Delta \lambda}$ and also transformations like \eqref{B1}.
Moreover,  we assume that the number of $\mathbf{k}$ points which are 
integrated out by use of the unitary transformation \eqref{G28} is small 
compared to the total number of $\mathbf{k}$ points. This 
assumption is needed for the evaluation of higher order 
commutators. For instance, the commutators which arise 
from repeated application of $X_ {\lambda, \Delta \lambda}$ to 
$( \hat{f}_{\mathbf{k}m}^{\dagger}c_{\mathbf{k}m}+\mathrm{h.c.})$ and 
$( f_{\mathbf{k}m}^{\dagger}c_{\mathbf{k}m}+\mathrm{h.c.})$
read $(n = 1, 2, 3, \cdots )$
\begin{eqnarray}
  \label{B24}
  \lefteqn{
    \mathbf{X}_{\lambda,\Delta\lambda}^{2n}
    \left( \hat{f}_{\mathbf{k}m}^{\dagger}c_{\mathbf{k}m}+\mathrm{h.c.}\right)
    \,=\,
  }&& \\
  &=&
  (-1)^{n} \, \Theta_{\mathbf{k}}(\lambda,\Delta\lambda)
  \left[
    2 \sqrt{D} A_{\mathbf{k}}(\lambda, \Delta\lambda) 
  \right]^{2n}
  \left\{
    \left( 
      \hat{f}_{\mathbf{k}m}^{\dagger}c_{\mathbf{k}m}+\mathrm{h.c.}
    \right)
    \phantom{\sum_{\tilde{m}(\not=m)}}
  \right. \nonumber\\[1ex]
  &&
  \left.
    \qquad\qquad +
    \frac{1}{2D}
    \left\langle
      \hat{f}_{\mathbf{k}m}^{\dagger}c_{\mathbf{k}m}+\mathrm{h.c.}
    \right\rangle
    \left[
      1 - D - 
      \sum_{\tilde{m}(\not=m)} 
      \left(
        \hat{f}_{\tilde{m}}^{\dagger} \hat{f}_{\tilde{m}}
      \right)_{\mathrm{L}}
    \right]
  \right\}, \nonumber \\[3ex]
  \label{B25}
  \lefteqn{
    \mathbf{X}_{\lambda,\Delta\lambda}^{2n+1}
    \left( \hat{f}_{\mathbf{k}m}^{\dagger}c_{\mathbf{k}m}+\mathrm{h.c.}\right)
    \,=\,
  } &&\\
  &=&
  \frac{(-1)^{n}}{\sqrt{D}} \, \Theta_{\mathbf{k}}(\lambda,\Delta\lambda)
  \left[ 
    2 \sqrt{D} A_{\mathbf{k}}(\lambda, \Delta\lambda)
  \right]^{2n+1}
  \left\{
    \left(
      \hat{f}_{\mathbf{k}m}^{\dagger} \hat{f}_{\mathbf{k}m}
    \right)_{\mathrm{NL}} +
    \left( \hat{f}_{m}^{\dagger} \hat{f}_{m} \right)_{\mathrm{L}}
    \phantom{\sum_{\tilde{m}(\not=m)}}
  \right.\nonumber\\[1ex]
  &&
  \left.
    \qquad\qquad -
    D c_{\mathbf{k}m}^{\dagger} c_{\mathbf{k}m} -
    \left\langle c_{\mathbf{k}m}^{\dagger} c_{\mathbf{k}m} \right\rangle -
    \left[
      1 - D - 
      \sum_{\tilde{m}(\not=m)} 
      \left(
        \hat{f}_{\tilde{m}}^{\dagger} \hat{f}_{\tilde{m}}
      \right)_{\mathrm{L}}
    \right]
  \right\}, \nonumber
\end{eqnarray}
\begin{eqnarray}
    \mathbf{X}_{\lambda,\Delta\lambda}^{2n}
    \left( f_{\mathbf{k}m}^{\dagger}c_{\mathbf{k}m}+\mathrm{h.c.}\right)
  &=&
  \mathbf{X}_{\lambda,\Delta\lambda}^{2n}
  \left( \hat{f}_{\mathbf{k}m}^{\dagger}c_{\mathbf{k}m}+\mathrm{h.c.}\right) +
  (-1)^{n} \, \Theta_{\mathbf{k}}(\lambda,\Delta\lambda)
  \left[ \sqrt{D} A_{\mathbf{k}}(\lambda, \Delta\lambda) \right]^{2n}
  \nonumber\\[1ex]
  &&
  \label{B26}
  \qquad\qquad \times
  \left\{
    \left( \hat{f}_{\mathbf{k}m}^{\dagger}c_{\mathbf{k}m}+\mathrm{h.c.}\right)
    -
    \left( f_{\mathbf{k}m}^{\dagger}c_{\mathbf{k}m}+\mathrm{h.c.}\right)
  \right\} \\[3ex]
  \mathbf{X}_{\lambda,\Delta\lambda}^{2n+1}
  \left( f_{\mathbf{k}m}^{\dagger}c_{\mathbf{k}m}+\mathrm{h.c.}\right)
  & = &
  \mathbf{X}_{\lambda,\Delta\lambda}^{2n+1}
  \left( \hat{f}_{\mathbf{k}m}^{\dagger}c_{\mathbf{k}m}+\mathrm{h.c.}\right) +
  \frac{(-1)^{n}}{\sqrt{D}} \Theta_{\mathbf{k}}(\lambda,\Delta\lambda)
  \left[ \sqrt{D} A_{\mathbf{k}}(\lambda, \Delta\lambda) \right]^{2n+1}
  \nonumber\\
  &&
  \label{B27}
  \qquad\qquad \times
  \left\{
    \left( 
      \hat{f}_{\mathbf{k}m}^{\dagger} f_{\mathbf{k}m} + \mathrm{h.c.}
    \right)_{\mathrm{NL}} - 
    2 \left( 
      \hat{f}_{\mathbf{k}m}^{\dagger} \hat{f}_{\mathbf{k}m}
    \right)_{\mathrm{NL}} 
  \right\}
\end{eqnarray}

To trace back all contributions to terms appearing in
the unperturbed Hamiltonian $\mathcal{H}_{0,\lambda}$, one has 
to replace all Hubbard operators   
by appropriate expressions in terms of usual Fermi operators. 
Thus, further approximations are needed for the local
$
  \left(
    \hat{f}_{\mathbf{k}m}^{\dagger} \hat{f}_{\mathbf{k}m}
  \right)_{\mathrm{L}}
$
and for the non-local $f$ electron particle-hole excitations
$
  \left(
    \hat{f}_{\mathbf{k}m}^{\dagger} \hat{f}_{\mathbf{k}m}
  \right)_{\mathrm{NL}} \ .
$
As was discussed 
before, due to the strong local Coulomb
interaction, only empty and singly occupied $f$ sites are of physical
relevance. Thus the operator
$
  \left(
    \hat{f}_{\mathbf{k}m}^{\dagger} \hat{f}_{\mathbf{k}m}
  \right)_{\mathrm{L}}
$
applied on physical states can not generate doubly occupied 
$f$ sites. Therefore, the operator can be replaced by
\begin{eqnarray}
  \label{B28}
  \left(
    \hat{f}_{\mathbf{k}m}^{\dagger} \hat{f}_{\mathbf{k}m}
  \right)_{\mathrm{L}}
  & \approx &
  \left(
    f_{\mathbf{k}m}^{\dagger} f_{\mathbf{k}m}
  \right)_{\mathrm{L}}.
\end{eqnarray}
The second operator, 
$
  \left(
    \hat{f}_{\mathbf{k}m}^{\dagger} \hat{f}_{\mathbf{k}m}
  \right)_{\mathrm{NL}}
$
represents an $f$ electron hopping between different sites without
creating doubly occupied $f$ sites. Thus, we may approximate
\begin{eqnarray}
  \label{B29}
  \left(
    \hat{f}_{\mathbf{k}m}^{\dagger} \hat{f}_{\mathbf{k}m} 
  \right)_{\mathrm{NL}}  
  & \approx &
  \frac{1}{N} \sum_{i,j(\neq i)} \hat{f}^{\dagger}_{im} f_{jm} \,
  e^{i{\bf k}({\bf R}_i -{\bf R}_j)} 
  \, = \,
  \frac{1}{N} \sum_{i,j(\neq i)} {\cal D}_{im} \, f^{\dagger}_{im} f_{jm} \,
  e^{i{\bf k}({\bf R}_i -{\bf R}_j)}  \\
  & \approx &
  D \left(
    f_{\mathbf{k}m}^{\dagger} f_{\mathbf{k}m} 
  \right)_{\mathrm{NL}}
  \nonumber
\end{eqnarray}
where in the last equation the creation of doubly occupied sites is only
fulfilled within a factorization approximation,  ${\cal D}_{im} \approx
D$.

Finally, by inserting \eqref{B24}-\eqref{B25} into \eqref{B1}, and by using the
approximations \eqref{B28} and \eqref{B29} one finds
\begin{eqnarray}
  \label{B30}
  \lefteqn{
    e^{X_{\lambda, \Delta \lambda}}
    c^{\dagger}_{{\bf k}m} c_{{\bf k}m} 
    e^{- X_{\lambda, \Delta \lambda}} -
    c^{\dagger}_{{\bf k}m} c_{{\bf k}m} 
    \,=\,
  }
  && \\[1ex]
  &=&
  - \,
  \frac{1}{2D} \, \Theta_{\mathbf{k}}(\lambda,\Delta\lambda) \,
  A_{\mathbf{k}}(\lambda, \Delta\lambda) \,
  \left\langle
    \hat{f}_{\mathbf{k}m}^{\dagger}c_{\mathbf{k}m}+\mathrm{h.c.}
  \right\rangle 
  \left[
    1 - D - 
    \sum_{\tilde{m}(\not=m)} 
    \left( f_{\tilde{m}}^{\dagger} f_{\tilde{m}} \right)_{\mathrm{L}}
  \right]
  \nonumber\\[1ex]
  &&
  -\,
  \frac{1}{2D} \, \Theta_{\mathbf{k}}(\lambda,\Delta\lambda) \,
  \left\{
    \cos \left[ 2\sqrt{D} A_{\mathbf{k}}(\lambda, \Delta\lambda) \right] - 
    1
  \right\}
  \left\{
    D \left( f_{\mathbf{k}m}^{\dagger} f_{\mathbf{k}m} \right)_{\mathrm{NL}} +
    \left( f_{m}^{\dagger} f_{m} \right)_{\mathrm{L}} 
    \phantom{\sum_{\tilde{m}(\not=m)}}
  \right. \nonumber \\[1ex]
  &&
  \left.
    \qquad\qquad -\,
    D c_{\mathbf{k}m}^{\dagger} c_{\mathbf{k}m} - 
    \left\langle c_{\mathbf{k}m}^{\dagger} c_{\mathbf{k}m} \right\rangle
    \left[
      1 - D - 
      \sum_{\tilde{m}(\not=m)} 
      \left( f_{\tilde{m}}^{\dagger} f_{\tilde{m}} \right)_{\mathrm{L}}
    \right]
  \right\} \nonumber \\[1ex]
  &&
  +\,
  \frac{1}{2\sqrt{D}} \, \Theta_{\mathbf{k}}(\lambda,\Delta\lambda) \,
  \sin \left[ 2\sqrt{D} A_{\mathbf{k}}(\lambda, \Delta\lambda) \right]
  \left\{
    \left( 
      \hat{f}_{\mathbf{k}m}^{\dagger}c_{\mathbf{k}m}+\mathrm{h.c.}
    \right) 
    \phantom{\sum_{\tilde{m}(\not=m)}}
  \right. \nonumber \\[1ex]
  &&
  \left.
    \qquad\qquad +\,
    \frac{1}{2D}
    \left\langle
      \hat{f}_{\mathbf{k}m}^{\dagger}c_{\mathbf{k}m}+\mathrm{h.c.}
    \right\rangle
    \left[
      1 - D - 
      \sum_{\tilde{m}(\not=m)} 
      \left( f_{\tilde{m}}^{\dagger} f_{\tilde{m}} \right)_{\mathrm{L}}
    \right]
  \right\}. \nonumber
\end{eqnarray}
Similar equations can be derived for the transformations of the
remaining operators 
\begin{eqnarray}
  \label{B31}
  \lefteqn{
    e^{X_{\lambda, \Delta \lambda}}
    \left( 
      f_{\mathbf{k}m}^{\dagger} f_{\mathbf{k}m} 
    \right)_{\mathrm{NL}} 
    e^{- X_{\lambda, \Delta \lambda}} -
    \left( 
      f_{\mathbf{k}m}^{\dagger} f_{\mathbf{k}m} 
    \right)_{\mathrm{NL}}
    \,=\,
  } &&\\ [1ex]
  &=&
  - \,D
  \left\{
    \left[
      e^{X_{\lambda, \Delta \lambda}}
      c^{\dagger}_{{\bf k}m} c_{{\bf k}m} 
      e^{- X_{\lambda, \Delta \lambda}} -
      c^{\dagger}_{{\bf k}m} c_{{\bf k}m}
    \right] + 
    \Theta_{\mathbf{k}}(\lambda,\Delta\lambda) \,
    A_{\mathbf{k}}(\lambda, \Delta\lambda) \,
    \left\langle
      \hat{f}_{\mathbf{k}m}^{\dagger}c_{\mathbf{k}m}+\mathrm{h.c.}
    \right\rangle
    \phantom{\sum_{\tilde{m}(\not=m)}}
  \right.
  \nonumber\\[1ex]
  &&
  \left.
    \qquad\qquad \times\,
    \left[
      1 - D - 
      \sum_{\tilde{m}(\not=m)} 
      \left( f_{\tilde{m}}^{\dagger} f_{\tilde{m}} \right)_{\mathrm{L}}
    \right]
  \right\} \nonumber \\[1ex]
  &&
  + \,
  \frac{D}{N} \sum_{\mathbf{k}'}
  \left\{
    \left[
      e^{X_{\lambda, \Delta \lambda}}
      c^{\dagger}_{{\bf k'}m} c_{{\bf k'}m} 
      e^{- X_{\lambda, \Delta \lambda}} -
      c^{\dagger}_{{\bf k'}m} c_{{\bf k'}m}
    \right] 
    \phantom{\sum_{\tilde{m}(\not=m)}}
  \right. \nonumber \\[1ex]
  &&
  \left.
    \qquad\qquad + \,
    \Theta_{\mathbf{k'}}(\lambda,\Delta\lambda) \,
    A_{\mathbf{k'}}(\lambda, \Delta\lambda) \,
    \left\langle
      \hat{f}_{\mathbf{k'}m}^{\dagger}c_{\mathbf{k'}m}+\mathrm{h.c.}
    \right\rangle
    \left[
      1 - D - 
      \sum_{\tilde{m}(\not=m)} 
      \left( f_{\tilde{m}}^{\dagger} f_{\tilde{m}} \right)_{\mathrm{L}}
    \right]
  \right\}, \nonumber
\end{eqnarray}
\begin{eqnarray}
  e^{X_{\lambda, \Delta \lambda}}
  \left( f_{m}^{\dagger} f_{m} \right)_{\mathrm{L}}
  e^{- X_{\lambda, \Delta \lambda}} - 
  \left( f_{m}^{\dagger} f_{m} \right)_{\mathrm{L}}
  &=&
  -\, \frac{1}{N} \sum_{\mathbf{k}'}
  \left[
    e^{X_{\lambda, \Delta \lambda}}
    c^{\dagger}_{{\bf k'}m} c_{{\bf k'}m} 
    e^{- X_{\lambda, \Delta \lambda}} -
    c^{\dagger}_{{\bf k'}m} c_{{\bf k'}m}
  \right] , \nonumber\\
  \label{B32} 
  && 
\end{eqnarray}
\begin{eqnarray}
  \label{B33}
  \lefteqn{
    e^{X_{\lambda, \Delta \lambda}}
    \left(
      \hat{f}_{\mathbf{k}'m}^{\dagger}c_{\mathbf{k}'m}+\mathrm{h.c.}
    \right)
    e^{- X_{\lambda, \Delta \lambda}} -
    \left(
      \hat{f}_{\mathbf{k}'m}^{\dagger}c_{\mathbf{k}'m}+\mathrm{h.c.}
    \right)
    \,=\,
  }
  &&\\[1ex]
  &=&
  \Theta_{\mathbf{k}}(\lambda,\Delta\lambda) 
  \left\{
    \cos \left[ 2\sqrt{D} A_{\mathbf{k}}(\lambda, \Delta\lambda) \right] - 1 
  \right\}
  \left\{
    \left( 
      \hat{f}_{\mathbf{k}m}^{\dagger}c_{\mathbf{k}m}+\mathrm{h.c.}
    \right) 
    \phantom{\sum_{\tilde{m}(\not=m)}}
  \right. \nonumber \\[1ex]
  &&
  \left.
    \qquad\qquad +\,
    \frac{1}{2D}
    \left\langle
      \hat{f}_{\mathbf{k}m}^{\dagger}c_{\mathbf{k}m}+\mathrm{h.c.}
    \right\rangle
    \left[
      1 - D - 
      \sum_{\tilde{m}(\not=m)} 
      \left( f_{\tilde{m}}^{\dagger} f_{\tilde{m}} \right)_{\mathrm{L}}
    \right]
  \right\} \nonumber \\[1ex]
  &&
  + \,
  \frac{1}{\sqrt{D}}
  \Theta_{\mathbf{k}}(\lambda,\Delta\lambda) 
  \sin \left[ 2\sqrt{D} A_{\mathbf{k}}(\lambda, \Delta\lambda) \right] 
  \left\{
    D \left( f_{\mathbf{k}m}^{\dagger} f_{\mathbf{k}m} \right)_{\mathrm{NL}} +
    \left( f_{m}^{\dagger} f_{m} \right)_{\mathrm{L}} 
    \phantom{\sum_{\tilde{m}(\not=m)}}
  \right. \nonumber \\[1ex]
  &&
  \left.
    \qquad\qquad -\,
    D c_{\mathbf{k}m}^{\dagger} c_{\mathbf{k}m} - 
    \left\langle c_{\mathbf{k}m}^{\dagger} c_{\mathbf{k}m} \right\rangle
    \left[
      1 - D - 
      \sum_{\tilde{m}(\not=m)} 
      \left( f_{\tilde{m}}^{\dagger} f_{\tilde{m}} \right)_{\mathrm{L}}
    \right]
  \right\}. \nonumber
\end{eqnarray}
where the second terms of the r.h.s. of Eqs.~\eqref{B26} and \eqref{B27} have
been neglected.

\section{Transformation of the one-particle operators}
\label{one-particle}
To determine the transformation of the one-particle operators
 we have again to apply the unitary transformation \eqref{G10}  which was
used before to renormalize the Hamiltonian. As in Ref.~\onlinecite{Becker} we 
first make the simplest operator ansatz for the $\lambda$ dependent 
$c$ creation operator
\begin{eqnarray}
  \label{C1}
  c_{\mathbf{k}m}^{\dagger}(\lambda) &=&
  u_{\mathbf{k},\lambda} c_{\mathbf{k}m}^{\dagger} + 
  v_{\mathbf{k},\lambda} \hat{f}_{\mathbf{k}m}^{\dagger}
\end{eqnarray}
with the initial parameter values corresponding to  
the unrenormalized operators 
($\lambda=\Lambda$)
\begin{eqnarray}
  \label{C2}
  u_{\mathbf{k},(\lambda=\Lambda)} &=& 1, \qquad
  v_{\mathbf{k},(\lambda=\Lambda)} \,=\, 0.
\end{eqnarray}
Because the $\lambda$ dependent operators have to fulfill the same
anticommutator relations as the unrenormalized operators, one concludes that
\begin{eqnarray}
  \label{C3}
  1 &=& 
  \left| u_{\mathbf{k},\lambda} \right|^{2} + 
  D \left| v_{\mathbf{k},\lambda} \right|^{2}
\end{eqnarray}
holds for all $\mathbf{k}$ and $\lambda$ values. Thus, the transformation 
of the $f$ electron creation operator is given by
\begin{eqnarray}
  \label{C4}
  \hat{f}_{\mathbf{k}m}^{\dagger}(\lambda) &=&
  - D \, v_{\mathbf{k},\lambda} c_{\mathbf{k}m}^{\dagger} + 
  u_{\mathbf{k},\lambda} \hat{f}_{\mathbf{k}m}^{\dagger}.
\end{eqnarray}
Thereby, the approximation 
$ [ \hat{f}_{\mathbf{k}m}^{\dagger}, \hat{f}_{\mathbf{k}m} ]_{+} \approx D $
was used.

To derive renormalization equations for the parameters
$u_{\mathbf{k},\lambda}$ and $v_{\mathbf{k},\lambda}$ of the one-particle
operators we again consider the transformation step from $\lambda$ to 
$(\lambda-\Delta\lambda)$. As in the case of the Hamiltonian [compare
Eqs.~\eqref{G31} and \eqref{G35}] we obtain two equations for
$c_{\mathbf{k}m}^{\dagger}(\lambda-\Delta\lambda)$
\begin{eqnarray}
  \label{C5}
  c_{\mathbf{k}m}^{\dagger}(\lambda-\Delta\lambda) &=&
  u_{\mathbf{k},(\lambda-\Delta\lambda)} c_{\mathbf{k}m}^{\dagger} + 
  v_{\mathbf{k},(\lambda-\Delta\lambda)} \hat{f}_{\mathbf{k}m}^{\dagger} \\
  \label{C6}
  &=&
  u_{\mathbf{k},\lambda} \,
  e^{X_{\lambda,\Delta\lambda}} 
  \, c_{\mathbf{k}m}^{\dagger} \, 
  e^{-X_{\lambda,\Delta\lambda}} 
  + 
  v_{\mathbf{k},\lambda} \,
  e^{X_{\lambda,\Delta\lambda}} 
  \, \hat{f}_{\mathbf{k}m}^{\dagger} \, 
  e^{-X_{\lambda,\Delta\lambda}} 
\end{eqnarray}
where the first one is derived from ansatz \eqref{C4}. The second 
equation follows
from the application of the unitary transformation \eqref{G11} to
$c_{\mathbf{k}m}^{\dagger}(\lambda)$. 
To calculate the transformed operators in Eq.~\eqref{C6} one has to retrace
the procedure of appendix \ref{Transformation} so that we obtain
\begin{eqnarray}
  \label{C7}
  e^{X_{\lambda,\Delta\lambda}} 
  \, c_{\mathbf{k}m}^{\dagger} \, 
  e^{-X_{\lambda,\Delta\lambda}}
  &=&
  c_{\mathbf{k}m}^{\dagger} + 
  \Theta_{\mathbf{k}}(\lambda,\Delta\lambda) 
  \left\{
    \cos \left[ \sqrt{D} A_{\mathbf{k}}(\lambda, \Delta\lambda) \right] - 1 
  \right\}
  c_{\mathbf{k}m}^{\dagger}  
  \\
  && +\,
  \frac{1}{\sqrt{D}}
  \Theta_{\mathbf{k}}(\lambda,\Delta\lambda) 
  \sin \left[ \sqrt{D} A_{\mathbf{k}}(\lambda, \Delta\lambda) \right] 
  \hat{f}_{\mathbf{k}m}^{\dagger}, \nonumber \\[2ex]
  \label{C8}
  e^{X_{\lambda,\Delta\lambda}} 
  \, \hat{f}_{\mathbf{k}m}^{\dagger} \, 
  e^{-X_{\lambda,\Delta\lambda}} 
  &=&
  \hat{f}_{\mathbf{k}m}^{\dagger} + 
  \Theta_{\mathbf{k}}(\lambda,\Delta\lambda) 
  \left\{
    \cos \left[ \sqrt{D} A_{\mathbf{k}}(\lambda, \Delta\lambda) \right] - 1 
  \right\}
  \hat{f}_{\mathbf{k}m}^{\dagger} \\
  && - \,
  \sqrt{D}
  \Theta_{\mathbf{k}}(\lambda,\Delta\lambda) 
  \sin \left[ \sqrt{D} A_{\mathbf{k}}(\lambda, \Delta\lambda) \right] 
  c_{\mathbf{k}m}^{\dagger}. \nonumber
\end{eqnarray}

Finally, inserting Eqs. \eqref{C7} and \eqref{C8} into \eqref{C6} we 
find the renormalization equations for the parameters 
$u_{\mathbf{k},\lambda}$ and $v_{\mathbf{k},\lambda}$, 
\begin{eqnarray}
  \label{C9}
  \lefteqn{
    u_{\mathbf{k},(\lambda-\Delta\lambda)} - u_{\mathbf{k},\lambda}
    \,=\,
  }\\
  &=&
  u_{\mathbf{k},\lambda} 
  \left\{ \cos 
    \left[ 
      \sqrt{D} A_{\mathbf{k}}(\lambda, \Delta\lambda) 
    \right] - 1 
  \right\}
  - \sqrt{D}\,v_{\mathbf{k},\lambda} \sin 
  \left[ \sqrt{D} A_{\mathbf{k}}(\lambda,\Delta\lambda) \right],
  \nonumber\\[1ex]
  \label{C10}
  \lefteqn{
    v_{\mathbf{k},(\lambda-\Delta\lambda)} - v_{\mathbf{k},\lambda}
    \,=\,
  }\\
  &=&
  v_{\mathbf{k},\lambda} 
  \left\{
    \cos \left[ \sqrt{D} A_{\mathbf{k}}(\lambda, \Delta\lambda) \right] - 1 
  \right\} + 
  \frac{1}{\sqrt{D}}
  \,u_{\mathbf{k},\lambda} 
  \sin \left[ \sqrt{D} A_{\mathbf{k}}(\lambda, \Delta\lambda) \right].
  \nonumber
\end{eqnarray}

\end{appendix}

\end{widetext}


\end{document}